\documentclass[journal]{IEEEtran}
\usepackage{amssymb}
\usepackage[cmex10]{amsmath}
\usepackage{graphicx}
\usepackage{subfigure}
\usepackage{url}
\usepackage{bm}
\usepackage{mathrsfs}
\usepackage{amsthm}
\usepackage{multirow}

\usepackage[colorlinks]{hyperref}
\usepackage[applemac]{inputenc}

\newtheorem{theorem}{Theorem}

\newtheorem{lemma}{Lemma}
\newtheorem{remark}{Remark}
\usepackage{algorithm}
\usepackage{algpseudocode}
\usepackage{tensor}
\usepackage{graphicx}
\usepackage{epstopdf}
\IEEEoverridecommandlockouts
	
\begin{document}

\title{Capacity Maximization for\\FAS-assisted Multiple Access Channels}

\author{
\IEEEauthorblockN{Hao Xu, \emph{Member, IEEE}\IEEEauthorrefmark{0},
		               Kai-Kit Wong, \emph{Fellow, IEEE}\IEEEauthorrefmark{0},
    		               Wee Kiat New, \emph{Member, IEEE}\IEEEauthorrefmark{0},\\
    		               Farshad Rostami Ghadi, \emph{Member, IEEE}\IEEEauthorrefmark{0},
		               Gui Zhou, \emph{Member, IEEE}\IEEEauthorrefmark{0},
		               Ross Murch, \emph{Fellow, IEEE}\IEEEauthorrefmark{0},\\
		               Chan-Byoung Chae, \emph{Fellow, IEEE}\IEEEauthorrefmark{0},
  		               Yongxu Zhu, \emph{Senior Member, IEEE}\IEEEauthorrefmark{0},
		               and Shi Jin, \emph{Fellow, IEEE}\IEEEauthorrefmark{0}
\vspace{-6mm}
	}
\thanks{The work of H. Xu, K. Wong, W. K. New, and F. Rostami Ghadi are supported by the European Union's Horizon 2020 Research and Innovation Programme under MSCA Grant No. 101024636, and the Engineering and Physical Sciences Research Council (EPSRC) under grant EP/W026813/1.}
\thanks{H. Xu, K. K. Wong, W. K. New, and F. Rostami Ghadi are with the Department of Electronic and Electrical Engineering, University College London, London WC1E7JE, United Kingdom. K. K. Wong is also affiliated with Yonsei Frontier Laboratory, Yonsei University, Seoul, 03722, Korea (e-mail: $\rm \{hao.xu, kai\text{-}kit.wong, a.new, f.rostamighadi \}@ucl.ac.uk$). G. Zhou is with the Institute for Digital Communications, Friedrich-Alexander-University Erlangen-N{\"u}rnberg (FAU), 91054 Erlangen, Germany (e-mail: $\rm gui.zhou@fau.de$). R. Murch is with the Department of Electronic and Computer Engineering and Institute for Advanced Study (IAS), Hong Kong University of Science and Technology, Clear Water Bay, Hong Kong SAR, China (e-mail: $\rm eermurch@ust.hk$). C. B. Chae is with the School of Integrated Technology, Yonsei University, Seoul, 03722, Korea (e-mail: $\rm cbchae@yonsei.ac.kr$). Y. Zhu is with the School of Engineering, University of Warwick, Coventry,	UK (e-mail: $\rm yongxu.zhu@warwick.ac.uk$). S. Jin is with the Frontiers Science Center for Mobile Information Communication and Security, Southeast University, Nanjing, China (e-mail: $\rm jinshi@seu.edu.cn$).
}
\thanks{Corresponding author: Kai-Kit Wong.}
}

\maketitle

\begin{abstract}
This paper investigates a multiuser millimeter-wave (mmWave) uplink system in which each user is equipped with a multi-antenna fluid antenna system (FAS) while the base station (BS) has multiple fixed-position antennas. Our primary objective is to maximize the system capacity by optimizing the transmit covariance matrices and the antenna position vectors of the users jointly. To gain insights, we start by deriving upper bounds and approximations for the capacity. Then we delve into the capacity maximization problem. Beginning with the simple scenario of a single user equipped with a single-antenna FAS, we demonstrate that a closed-form optimal solution exists when there are only two propagation paths between the user and the BS. In the case where multiple propagation paths are present, a near-optimal solution can also be obtained through a one-dimensional search method. Expanding our focus to multiuser cases, in which users are equipped with either single- or multi-antenna FAS, we show that the original capacity maximization problems can be reformulated into distinct rank-one programmings. Then, we propose alternating optimization algorithms to deal with the transformed problems. Simulation results indicate that FAS can improve the capacity of the multiple access channel (MAC) greatly, and the proposed algorithms outperform all the benchmarks.
\end{abstract}

\begin{IEEEkeywords}
Fluid antenna system (FAS), multiple access, millimeter-wave communication, capacity maximization.
\end{IEEEkeywords}

\IEEEpeerreviewmaketitle

\section{Introduction}\label{section1}
\IEEEPARstart{R}{ecently}, a new promising technology, widely known as fluid antenna system (FAS), has been proposed for the sixth generation (6G) mobile communication systems, e.g., \cite{wong-ell2020,wong2022bruce,wong2023fluid,new2024tutorial}. By adjusting the antenna position within a spatial region, a transceiver gains the ability to navigate through the fluctuations in a fading channel, thereby providing additional degrees of freedom and significant performance gains \cite{new2023information}. The concept of FAS is not limited to a specific implementation. It can be implemented using the surface-wave based technology \cite{Shen-fas2021} or reconfigurable pixel technology \cite{song2013efficient}. The former requires a soft-material radiator to be mobilized in a confined space using nano-pumps, while the latter is composed of a matrix of pixels connected with electronic switches \cite{song2013efficient}. By turning the pixels on or off, the structure can make an antenna instantly appear or disappear in a given space. Experimental results on FAS have recently been reported in \cite{shen2024design} and \cite{zhang2024pixel}.

A large body of literature has studied the performance of FAS in point-to-point systems in terms of outage probability \cite{9131873,wong2021fluid, Tlebaldiyeva-2022, ghadi2023copula}, diversity gain \cite{psomas2023diversity, new2023fluid}, data rate \cite{new2023lett}, secrecy rate \cite{tang2023fluid, xu2024coding}, and etc. Specifically, FAS was first proposed by Wong {\em et al.}~in \cite{9131873,wong2021fluid}, and the effect of its size and resolution on the probability was examined. The analysis was subsequently extended to Nakagami fading channels by \cite{Tlebaldiyeva-2022}. By exploiting copula theory, \cite{ghadi2023copula} derived a closed-form expression of the outage probability under arbitrary correlated fading for a FAS-assisted point-to-point channel. More recently in \cite{psomas2023diversity}, three FAS architectures were considered, and a linear prediction scheme as well as space-time coded modulations were devised to enhance diversity. Adopting the channel model in \cite{khammassi2023new}, which can more accurately characterize the spatial correlation among the FAS ports, \cite{new2023fluid} studied both the outage probability and diversity gain of FAS. In \cite{new2023lett} and \cite{tang2023fluid}, the rate and secrecy rate of FAS-aided systems were respectively studied. Then it was shown by \cite{xu2024coding} that, using the coding-enhanced jamming strategy proposed by \cite{xu2022new}, the secrecy rate of the FAS-aided system in \cite{tang2023fluid} can be further increased.

The performance of using FAS in supporting multiple access channel (MAC) communications has also been explored, and this technique is called fluid antenna multiple access (FAMA) \cite{wong2022fluid, Wong-ELL2022a, wong2023slow, xu2024revisiting}. The idea lies in the fact that multiuser signals fade independently in space and as such, the FAS at a given user can operate at the port where the interfering users all fade deeply to have interference-less signal reception for communication. Depending on how fast the user updates the port, FAMA can be classified into fast \cite{wong2022fluid,Wong-ELL2022a} and slow FAMA \cite{wong2023slow}. Most recently in \cite{xu2024revisiting}, the outage probability of a two-user FAMA system was analyzed based on approximated channel models that characterize the space correlation among ports in a simplified but still accurate way.

All the above references, nevertheless, focus on the sub-6G frequency band, assuming rich scattering channels. To meet the ultra-high communication performance requirements, the use of millimeter-wave (mmWave) technology is essential for 6G communication systems. Therefore, it is crucial to investigate the performance of FAS subject to the channel characteristics appropriate for the mmWave band. Several papers have already explored this topic \cite{chen2024joint, ye2023fluid, xu2023channel, ma2023mimo}. Specifically, \cite{chen2024joint} maximized the energy efficiency of a point-to-point system in near-field communications and confirmed the effectiveness of FAS. Since it is not easy to obtain the instantaneous channel state information (CSI), in \cite{ye2023fluid}, the rate of a FAS-assisted point-to-point system was maximized based on the statistical CSI. Also, in \cite{xu2023channel}, a low-sample size method was proposed to estimate and reconstruct the CSI of a mmWave MAC system, where all transmitters use FAS. Later in \cite{ma2023mimo}, the capacity of a point-to-point mmWave system, in which both the transmitter and receiver sides are equipped with a multi-antenna FAS each, was investigated.\footnote{While the work in \cite{ma2023mimo} is under the name `movable antenna system', it is worth pointing out that FAS includes all forms of movable and non-movable flexible-position antenna systems. In other words, FAS represents any flexible-position antenna system irrespective of the implementation.} However, thus far, the capacity performance of FAS in multiuser systems, and how to achieve the capacity gains of FAS, are not well understood.

This paper considers a multiuser mmWave uplink system, where each user is equipped with a multi-antenna FAS while the BS has multiple fixed-position antennas. Our objective is to maximize the sum capacity of the system by optimizing the transmit covariance matrices and antenna position vectors of the users. The main contributions are summarized below.
\begin{itemize}
\item Given the channel characteristics of the mmWave band, we first derive an upper bound on the maximum system capacity. Based on this bound, we show that if there is only one propagation path, i.e., line-of-sight (LoS) path, between a user and the BS, in order to maximize capacity, this user should transmit at the maximum power and align its beamforming vector with the steering vector. In this case, adjusting the positions of this user's antennas will not provide any capacity gain. If all users have only a LoS path between them and the BS, the derived upper bound is tight, and a closed-form expression for the maximum system capacity can be obtained. In addition, we prove that if all users have many antennas, the maximum system capacity approaches to that of a fixed-antenna system.

\item We address the capacity maximization problem by considering different cases. First, we consider the simple scenario with a single-antenna user and show that, if there are two propagation paths between the user and the BS, a closed-form optimal solution is possible. If there are multiple propagation paths, a near-optimal solution can be found by using the one-dimensional search method. Next, we consider the multiuser case in which each user has a single antenna. In this case, all users should transmit at the maximum power and we only need to optimize their antenna positions. We provide two algorithms to solve the problem. The first is a straightforward one that iteratively optimizes the position of each user's antenna. The scheme in the single-user case can thus be applied in each iteration. In the second algorithm, we transform the original problem into a rank-one programming. Afterwards, we propose an algorithm to jointly optimize the positions of all users' antennas. Once the rank-one problem is solved, we map its solution to that of the original problem. Finally, we consider the general multiuser multi-antenna scenario. Although the problem becomes much more complicated, we show that it can also be transformed to the rank-one problem and thus be efficiently solved.

\item Simulation results show that FAS can greatly improve the capacity of the considered MAC system, and the proposed algorithms outperform the benchmarks. Furthermore, we undertake validation by simulation to confirm the connections between the system capacity, the established upper bound, and the derived approximations.
\end{itemize}

The rest of this paper is organized as follows. In Section~\ref{section2}, the system model is provided and the problem is formulated. In Section~\ref{section3}, we provide an upper bound and analyze the system capacity for some special cases. In Sections~\ref{single_an_FAS} and \ref{multi_an_FAS}, we solve the capacity maximization problems for the single- and multi-antenna FAS cases, respectively. Simulation results are given in Section~\ref{simul}, and we conclude the paper in Section~\ref{conclusion}.

Notations: $\mathbb R$, $\mathbb R_+$, and $\mathbb C$, respectively, represent the real, non-negative real, and complex spaces. Moreover, boldface lower and upper case letters are used to denote vectors and matrices. ${\bm I}_N$ stands for the $N \times N$ dimensional identity matrix and $\bm 0$ denotes the all-zero vector or matrix. Also, superscripts $(\cdot)^T$ and $(\cdot)^H$, respectively, denote the transpose and conjugate transpose operations. $\left\| \cdot \right\|_F$, $ \left\| \cdot \right\|_*$, and $\left\| \cdot \right\|_2$ are respectively the Frobenius, nuclear, and spectral norms of a matrix, and $\langle \cdot, \cdot \rangle$ represents the inner product of two vectors. Unless otherwise specified, the logarithm function $\log$ is of base $2$.

Before going into the main part, we would like to present some equations and inequalities of matrices, since they will be used frequently in this paper. First, for any matrices $\bm O_1$ and $\bm O_2$, if the dimensions match, we have \cite{petersen2008matrix}
\begin{align}
| \bm O_1 \bm O_2 | & = | \bm O_1 | |\bm O_2 |, \label{O1O2_0}\\
| \bm O_1 \bm O_2 + \bm I | & = | \bm O_2 \bm O_1 + \bm I |, \label{O1O2_1}\\
{\text {tr}} (\bm O_1 \bm O_2) & = {\text {tr}} (\bm O_2 \bm O_1), \label{O1O2_2}
\end{align}
where $\bm I$ is an identity matrix. 
Next, for positive semi-definite (PSD) matrices $\bm O_3, \bm O_4  \in {\mathbb C}^{N \times N}$, we have
\begin{equation}\label{O3O4}
{\text {tr}} (\bm O_3 \bm O_4) \leq {\text {tr}} (\bm O_3) {\text {tr}} (\bm O_4),
\end{equation}
which holds with equality if and only if $\bm O_3$ and $\bm O_4$ satisfy $\bm O_3 = \bm o \bm o^H$ and  $\bm O_4 = \eta \bm o \bm o^H$, in which $\eta \in {\mathbb R}_+$ is a non-negative real constant and $\bm o \in {\mathbb C}^{N \times 1}$. Note that the inequality (\ref{O3O4}) is proven by \cite[Theorem~1]{coope1994matrix}. However, the condition under which (\ref{O3O4}) holds with equality is not given in \cite{coope1994matrix} or anywhere else. Since it is important for our analysis, we prove the above condition in Appendix~\ref{equal_cond}.

\section{System Model}\label{section2}

\begin{figure}[]
\centering
\includegraphics[scale=0.4]{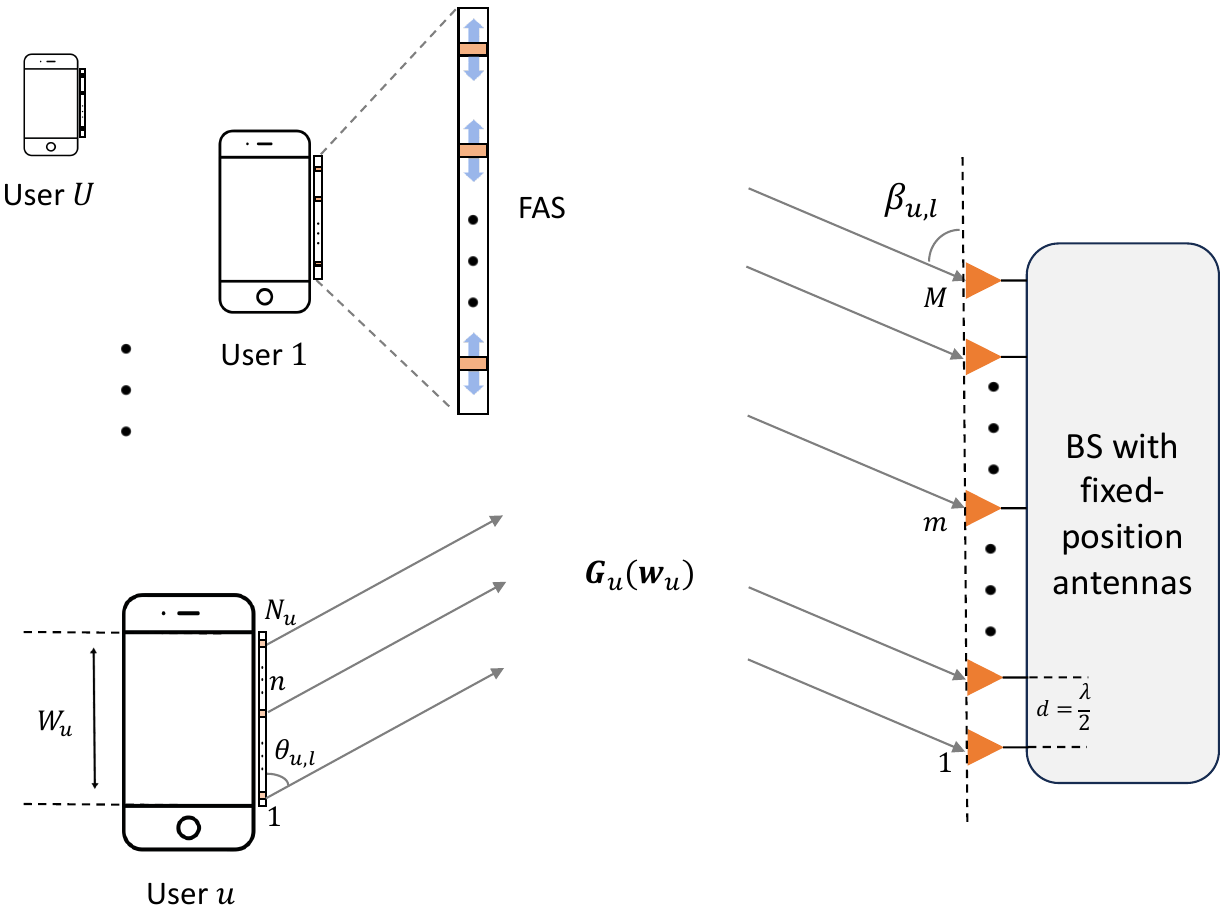}
\caption{A FAS-assisted uplink MAC where the users have FAS for transmission but the BS uses multiple fixed-position antennas for reception.}\label{system_model}
\vspace{-1 em}
\end{figure}

\subsection{Signal Model}
As depicted in Fig.~\ref{system_model}, we consider a narrow-band mmWave uplink system in which $U$ users simultaneously communicate with the BS using the same time-frequency resource. The BS is equipped with an $M$-antenna uniform linear array (ULA) with antenna spacing $d = \frac{\lambda}{2}$, where $\lambda$ is the wavelength. Each user is equipped with an $N_u$-antenna linear FAS of size $W_u$, and each antenna can be instantly switched to any position in a FAS.\footnote{Each antenna structure can be seen as a reconfigurable pixel-based linear FAS that has many tiny pixels, among which certain number of pixels and their connections are activated at each time to switch between many reconfigurable states \cite{zhang2024pixel}. Using this technology, it is possible to switch the antennas with almost no time delay. Also, extension to a two-dimensional (2D) FAS surface at each user is possible but at the expense of complication in the notations. We opt for the one-dimensional FAS case for simplicity.} Let $w_{u, n} \in [0, W_u]$ denote the position of the $n$-th antenna of user $u$ and $\bm w_u = [w_{u, 1}, \dots, w_{u, N_u}]^T$. 
Note that the antennas on the same FAS should be in different positions. Then the received signal at the BS is given by
\begin{equation}\label{yx}
\bm y = \sum_{u = 1}^U \bm G_u (\bm w_u) \bm x_u + \bm z,
\end{equation}
where $\bm x_u \sim {\cal {CN}} (\bm 0, \bm Q_u)$ denotes the signal vector of user~$u$, $\bm Q_u \in {\mathbb C}^{N_u \times N_u}$ is the transmit covariance matrix, $\bm G_u (\bm w_u) \in {\mathbb C}^{M \times N_u}$ denotes the channel matrix from user $u$ to the BS, and $\bm z \sim {\cal {CN}} (\bm 0, \bm I_M)$ represents the additive white Gaussian noise (AWGN). Then, the sum capacity of the system is given by \cite[Chapter~9.2.1]{el2011network}
\begin{equation}\label{capacity}
C(\bm Q_{\cal U}, \bm w_{\cal U}) = \log \left| \sum_{u=1}^U \bm G_u (\bm w_u) \bm Q_u \bm G_u^H (\bm w_u) + \bm I_M \right|,
\end{equation}
where ${\cal U} = \{1, \dots, U\}$, $\bm Q_{\cal U} = \{\bm Q_1, \dots, \bm Q_U\}$, and $\bm w_{\cal U} = \{\bm w_1, \dots, \bm w_U\}$. Also, we define ${\cal N}_u=\{1,\dots,N_u\}$.

\subsection{Channel Model}
Using the planar-wave geometric channel model typical in mmWave systems \cite{akdeniz2014millimeter}, $\bm G_u (\bm w_u)$ can be modeled as
\begin{equation}\label{G_u}
\bm G_u (\bm w_u) = \sqrt{M N_u} \sum_{l = 1}^{L_u} \gamma_{u,l} \bm a_{u, {\text R}} (\beta_{u,l}) \bm a_{u, {\text T}}^H (\theta_{u, l}, \bm w_u),
\end{equation}
where $L_u$ is the number of propagation paths between user $u$ and the BS, and $\gamma_{u,l}$ is the complex channel gain of the $l$-th path. Also, $\bm a_{u, {\text R}} (\beta_{u, l})$ and $\bm a_{u, {\text T}} (\theta_{u, l}, \bm w_u)$ are, respectively, the steering vectors at the receiver and transmitter sides, given by
\begin{align}\label{a_RT}
	& \bm a_{u, {\text R}} (\beta_{u, l})\notag\\
	& = \frac{1}{\sqrt{\!M}} \!\left[\! 1, e^{\!-\! j \frac{2 \pi}{\lambda} d \cos \beta_{u, l}}, \dots, e^{\!-\! j \frac{2 \pi}{\lambda} (M\!-\!1) d \cos \beta_{u, l}} \!\right]^{\!T},\\
	& \bm a_{u, {\text T}} (\theta_{u, l}, \bm w_u) \nonumber\\
	& = \frac{1}{\sqrt{N_u}} \left[e^{- j \frac{2 \pi}{\lambda} w_{u, 1} \cos \theta_{u, l}}, \dots, e^{- j \frac{2 \pi}{\lambda} w_{u, N_u} \cos \theta_{u, l}} \right]^T,
\end{align}
where $\beta_{u, l}, \theta_{u, l} \in [0, \pi]$ are, respectively, the angle-of-arrival (AoA) and angle-of-departure (AoD) of the $l$-th path of user~$u$. Now, we define
\begin{align}\label{A_RT}
\bm \varGamma_u &= {\rm diag} \{ \gamma_{u,1}, \dots, \gamma_{u, L_u} \} \in {\mathbb C}^{L_u \times L_u}, \nonumber\\
\bm A_{u,{\text R}} &= \left[ \bm a_{u,{\text R}} (\beta_{u, 1}), \dots, \bm a_{u,{\text R}} (\beta_{u, L_u}) \right] \in {\mathbb C}^{M \times L_u}, \nonumber\\
\bm A_{u,{\text T}} (\bm w_u) &= \left[ \bm a_{u,{\text T}} (\theta_{u, 1}, \bm w_u), \dots, \bm a_{u,{\text T}} (\theta_{u, L_u}, \bm w_u) \right] \!\in\! {\mathbb C}^{N_u \!\times\! L_u},
\end{align}
based on which the channel matrix $\bm G_u (\bm w_u)$ in (\ref{G_u}) can also be expressed as
\begin{equation}\label{G_u2}
\bm G_u (\bm w_u) = \sqrt{M N_u} \bm A_{u,{\text R}} \bm \varGamma_u \bm A_{u,{\text T}}^H (\bm w_u).
\end{equation}
In this paper, our emphasis is on how the optimization can be done and we assume that the BS knows the CSI of all the users when the optimization is conducted.

\subsection{Problem Formulation}
We aim to maximize the system capacity by optimizing the transmit covariance matrices and antenna position vectors of all users. The problem can be formulated as
\begin{align}\label{problem1}
	\mathop {\max }\limits_{\bm Q_{\cal U}, \bm w_{\cal U}} \quad & C(\bm Q_{\cal U}, \bm w_{\cal U}) \nonumber\\
	\text{s.t.} \quad\;\; & \bm Q_u \succeq \bm 0, ~ {\text {tr}} (\bm Q_u) \leq P_u, ~\forall u \in {\cal U}, \nonumber\\
	& 0 \leq w_{u, n} \leq W_u, ~\forall u \in {\cal U},~ n \in {\cal N}_u, \nonumber\\
	& w_{u, n} \neq w_{u, n'}, ~\forall u \in {\cal U},~ n, n' \in {\cal N}_u,~ n \neq n',
\end{align}
where $P_u$ denotes the maximum transmit power of user~$u$, and the last constraint restricts that no two antennas of the same FAS should be in the same position.

It should be noted that capacity maximization by precoding or beamforming is a classical problem in multiple-input multiple-output (MIMO) MAC communication systems \cite{yu2004iterative, goldsmith2003capacity, kobayashi2006iterative}. For MIMO MAC, where all nodes use traditional fixed-position antennas, the problem only optimizes the transmit covariance matrices, is convex, and can therefore be optimally solved by the well-known iterative water-filling method \cite[Chapter~9.2.1]{el2011network}. Nevertheless, in our case, where each user has multiple fluid antennas, the problem becomes much more complicated. This is because, due to the mobility property of the fluid antennas, the channel matrices $\bm G_u (\bm w_u), \forall u \in {\cal U}$ are no longer fixed, but reconfigurable, making the problem highly non-convex. In the following sections, we first analyze (\ref{problem1}) and then propose several alternative algorithms to solve it.

\section{Upper Bound and Approximation}\label{section3}
As (\ref{problem1}) is intractable, we first derive an upper bound on the maximum system capacity, and then consider some special or extreme cases, in which a closed-form expression or approximation of the maximum system capacity can be derived. These results not only provide valuable insights, but also serve as benchmarks for evaluating the capacity performance of FAS.

\begin{theorem}\label{theorem0}
The sum capacity of the considered system is upper bounded by the optimal objective function value of the following convex problem:
\begin{align}\label{problem1_1}
\mathop {\max }\limits_{{\tilde{\bm Q}}_{\cal U}} \quad & C^{\rm ub} ({\tilde{\bm Q}}_{\cal U}) \nonumber\\
{\rm s.t.} \quad\; & {\tilde{\bm Q}}_u \succeq \bm 0, ~ {\rm tr} ({\tilde{\bm Q}}_u) \leq L_u P_u, ~\forall u \in {\cal U},
\end{align}
where ${\tilde{\bm Q}}_u \in {\mathbb C}^{L_u \times L_u}$, ${\tilde{\bm Q}}_{\cal U} = \{{\tilde{\bm Q}}_1, \dots, {\tilde{\bm Q}}_U\}$, and 
\begin{equation}\label{capacity_ub}
C^{\rm ub} ({\tilde{\bm Q}}_{\cal U}) \!=\! \log \left| \sum_{u=1}^U M N_u \bm A_{u,{\text R}} \bm \varGamma_u {\tilde{\bm Q}}_u \bm \varGamma_u^H \bm A_{u,{\text R}}^H + \bm I_M \right|.\!\!
\end{equation}
\end{theorem}

\itshape \textbf{Proof:}  \upshape
See Appendix~\ref{prove_theorem0}.
\hfill $\Box$

Obviously, (\ref{problem1_1}) can be seen as the sum capacity maximization problem of a $U$-user uplink system with channel matrix $\sqrt{M N_u} \bm A_{u,{\text R}} \bm \varGamma_u$ between user~$u$ and the BS. This problem is convex and its optimal solution can be efficiently obtained by using the iterative water-filling method \cite[Chapter~9]{el2011network}.\footnote{Water-filling and iterative water-filling methods are frequently mentioned in this paper. Due to space limitations and the fact that they are standard and well known, we will not provide the details here. One may refer to \cite{el2011network} for more details about these methods.} Although, as we will show by computer simulation results, this bound is generally loose, we prove below that it is tight in some cases and can greatly simplify problem (\ref{problem1}).

\begin{lemma}\label{lemma0}
For any user $u \in {\cal U}$, if $L_u = 1$, i.e., there is only one path between this user and the BS, then problem (\ref{problem1}) is equivalent to
\begin{align}\label{problem1_2}
\mathop {\max}_{\bm Q_{{\cal U}'}, \bm w_{{\cal U}'}}  & \log \left|\begin{array}{l}
\sum_{u' \in {\cal U}'} \bm G_{u'} (\bm w_{u'}) \bm Q_{u'} \bm G_{u'}^H (\bm w_{u'})\\
+ M N_u P_u |\gamma_{u,1}|^2 \bm a_{u,{\text R}} (\beta_{u, 1}) \bm a_{u,{\text R}}^H (\beta_{u, 1}) + \bm I_M
\end{array}\right| \nonumber\\
{\rm s.t.}~& \bm Q_{u'} \succeq \bm 0, ~ {\rm tr} (\bm Q_{u'}) \leq P_{u'}, ~\forall u' \in {\cal U}', \nonumber\\
& 0 \leq w_{u', n} \leq W_{u'}, ~\forall u' \in {\cal U}',~ n \in {\cal N}_{u'}, \nonumber\\
& w_{u', n} \neq w_{u', n'}, ~\forall u' \in {\cal U}',~ n, n' \in {\cal N}_{u'},~ n \neq n'\!,\!\!\!
\end{align}
where ${\cal U}' = {\cal U} \setminus \{u\}$.
\end{lemma}

\itshape \textbf{Proof:}  \upshape
See Appendix~\ref{prove_lemma0}.
\hfill $\Box$

Lemma~\ref{lemma0} together with its proof in Appendix~\ref{prove_lemma0} shows that if there is only one path between user~$u$ and the BS, we only need to optimize the covariance matrices and antenna position vectors of the other users. As for user~$u$, it should transmit at the maximum power and align its beamforming vector with the steering vector $\bm a_{u,{\text T}} (\theta_{u, 1}, \bm w_u)$ (see (\ref{Q_u})). Adjusting the positions of this user's antennas does not bring any gain.

\begin{lemma}\label{lemma01}
If $L_u = 1, \forall u \in {\cal U}$, i.e., there is only one path between any user and the BS, then in the optimal case, the upper bound provided in Theorem~\ref{theorem0} is tight and
\begin{align}\label{tight_opt}
C^*& = C^{\rm ub*} \nonumber\\
& = \log \left| \sum_{u = 1}^U M N_u P_u |\gamma_{u,1}|^2 \bm a_{u,{\text R}} (\beta_{u, 1}) \bm a_{u,{\text R}}^H (\beta_{u, 1}) \!+\! \bm I_M \right|.\!\!\!
\end{align}
\end{lemma}

\itshape \textbf{Proof:}  \upshape
If $L_u = 1, \forall u \in {\cal U}$, then it is apparant from Lemma~\ref{lemma0} that the optimal objective function value of (\ref{problem1}), i.e., $C^*$, is equal to (\ref{tight_opt}). Moreover, in this case, ${\tilde{\bm Q}}_u$ in (\ref{problem1_1}) is a scalar variable within the interval $[0, P_u]$. Since $C^{\text {ub}} ({\tilde{\bm Q}}_{\cal U})$ monotonically increases with ${\tilde{\bm Q}}_u, \forall u \in {\cal U}$, all users should transmit at the maximum power in the optimal case. Therefore, $C^{\text {ub} *}$ is also equal to (\ref{tight_opt}), which completes the proof.
\hfill $\Box$

\begin{remark}
	Interestingly, the observations made in Lemma~\ref{lemma0} and Lemma~\ref{lemma01} can also be explained by the condition under which (\ref{O3O4}) holds with equality.
	In particular, if $L_u = 1$, we know from (\ref{A_RT}) that $\bm A_{u,{\text T}} (\bm w_u) = \bm a_{u,{\text T}} (\theta_{u, 1}, \bm w_u)$. 
	Then, $\bm A_{u,{\text T}} (\bm w_u) \bm A_{u,{\text T}} (\bm w_u)^H = \bm a_{u,{\text T}} (\theta_{u, 1}, \bm w_u) \bm a_{u,{\text T}} (\theta_{u, 1}, \bm w_u)^H$ is a rank-one matrix.
	From the condition provided after (\ref{O3O4}) we know that the first inequality in (\ref{trace_ub}), i.e.,
	\begin{equation}\label{ineq}
		{\text {tr}} (\bm A_{u,{\text T}} (\bm w_u) \bm A_{u,{\text T}}^H (\bm w_u) \bm Q_u) \leq {\text {tr}} (\bm A_{u,{\text T}} (\bm w_u) \bm A_{u,{\text T}}^H (\bm w_u)) {\text {tr}} (\bm Q_u),
	\end{equation}
	holds with equality if and only if $\bm Q_u$ takes on the form $\bm Q_u = \eta_u \bm a_{u,{\text T}} (\theta_{u, 1}, \bm w_u) \bm a_{u,{\text T}} (\theta_{u, 1}, \bm w_u)^H$, where $\eta_u$ is a non-negative real constant.
	Note that to maximize the system capacity, $\eta_u$ should be $P_u$.
	Then, $\bm Q_u = P_u \bm a_{u,{\text T}} (\theta_{u, 1}, \bm w_u) \bm a_{u,{\text T}} (\theta_{u, 1}, \bm w_u)^H$, which coincides with (\ref{Q_u}), and both inequalities in (\ref{trace_ub}) hold with equality.
	The upper bound provided in Theorem~\ref{theorem0} is thus tight if $L_u = 1, \forall u \in {\cal U}$. \hfill $\lozenge$
\end{remark}

\begin{theorem}\label{theorem3}
If $N_u, \forall u \in {\cal U}$ are all large, the maximum system capacity can be approximated by the optimal objective function value of the following convex problem:
\begin{align}\label{problem_approx}
\mathop {\max }_{{\hat {\bm Q}}_{\cal U}}~& C^{\rm approx} ({\hat {\bm Q}}_{\cal U}) \nonumber\\
{\rm s.t.}~& {\hat {\bm Q}}_u \succeq \bm 0, ~ {\rm tr} ({\hat {\bm Q}}_u) \leq P_u, ~\forall u \in {\cal U}, 
\end{align}
where ${\hat {\bm Q}}_u \in {\mathbb C}^{L_u \times L_u}$, ${\hat {\bm Q}}_{\cal U} = \{ {\hat {\bm Q}}_1, \dots, {\hat {\bm Q}}_U \}$, and 
\begin{align}\label{approx5}
C^{\rm approx} ({\hat {\bm Q}}_{\cal U}) \!=\! \log \left| \sum_{u=1}^U \!M N_u \bm A_{u,{\text R}} \bm \varGamma_u {\hat {\bm Q}}_u \bm \varGamma_u^H\! \bm A_{u,{\text R}}^H \!+\! \bm I_M \right|\!.\!\!\!
\end{align}
\end{theorem}

\itshape \textbf{Proof:}  \upshape
See Appendix~\ref{prove_theorem3}.
\hfill $\Box$

Similar to (\ref{problem1_1}), problem (\ref{problem_approx}) can also be optimally solved by the iterative water-filling method.
Theorem~\ref{theorem3} indicates that when the users have many antennas, the positions of their antennas barely affect the system capacity, and the maximum sum capacity can be approximated by the solution of (\ref{problem_approx}).

\begin{remark}
It should be noted that although the formulas in Theorem~\ref{theorem0} and Theorem~\ref{theorem3} look similar, they are fundamentally different. The covariance matrix ${\tilde{\bm Q}}_u$ in (\ref{problem1_1}) has power limit $L_u P_u$ and Theorem~\ref{theorem0}  is always true in any case. In contrast, ${\hat {\bm Q}}_u$ in (\ref{problem_approx}) has power limit $P_u$ and Theorem~\ref{theorem3} holds only when $N_u, \forall u \in {\cal U}$ are all large. \hfill $\lozenge$
\end{remark}

\begin{table*}[ht]
	\caption{Different scenarios and corresponding solutions to problem (\ref{problem1})}\label{table1}
	\begin{center}
	\begin{tabular}{|ccc|c|}
		\hline
		\multicolumn{3}{|c|}{\textbf{Scenario}}                                                &       \textbf{Solution of (\ref{problem1})}          \\ \hline
		\multicolumn{3}{|c|}{Single-path, $L_u = 1, \forall u \in {\cal U}$ }                                                              &      The optimal solution in closed-form is given in Lemma~\ref{lemma01}          \\ \hline
		\multicolumn{1}{|c|}{\multirow{4}{*}{Single-antenna FAS, $N_u = 1, \forall u \in {\cal U}$}} & \multicolumn{1}{c|}{\multirow{2}{*}{Single user}} & $L = 2$ &     The optimal solution in closed-form is given in Theorem~\ref{theorem1}              \\ \cline{3-4} 
		\multicolumn{1}{|c|}{}                  & \multicolumn{1}{c|}{}                  &  $L > 2$ &   A near-optimal solution can be found based on (\ref{find_w})              \\ \cline{2-4} 
		\multicolumn{1}{|c|}{}                  & \multicolumn{2}{c|}{\multirow{2}{*}{Multiple users}}    & \multirow{2}{*}{Algorithm~\ref{algorithm1} or Algorithm~\ref{algorithm2}} \\
		\multicolumn{1}{|c|}{}                  & \multicolumn{2}{c|}{}                     &                   \\ \hline
		\multicolumn{1}{|c|}{\multirow{2}{*}{Multi-antenna FAS}} & \multicolumn{2}{c|}{Single user}                     &      Algorithm~\ref{algorithm3}             \\ \cline{2-4} 
		\multicolumn{1}{|c|}{}                  & \multicolumn{2}{c|}{Multiple users}                     &       Algorithm~\ref{algorithm4}            \\ \hline
	\end{tabular}
\end{center}
\end{table*}

\section{Single-Antenna FAS}\label{single_an_FAS}
Here and in the next section, we start by on solving (\ref{problem1}) under simple scenarios. As we will show below that in some cases, the optimal closed-form solution of (\ref{problem1}) can be found, but in other cases finding the optimal solution is impossible. Thus, we propose specific methods targeting a particular set of circumstances. All considered scenarios and the corresponding solutions to problem (\ref{problem1}) are listed in Table~\ref{table1}.

First, we consider the case where each FAS has only one antenna in this section. For brevity, we omit the antenna index `$n$'. In this case, $\bm Q_u$ and $\bm w_u$ reduce to $q_u$ and $w_u$, and the channel matrix in (\ref{G_u}) or (\ref{G_u2}) becomes a vector
\begin{align}\label{G_u3}
\bm g_u (w_u) & = \sqrt{M} \sum_{l = 1}^{L_u} \gamma_{u,l} \bm a_{u, {\text R}} (\beta_{u,l}) a_{u, {\text T}}^* (\theta_{u, l}, w_u) \nonumber\\
& = \sqrt{M} \bm A_{u,{\text R}} \bm \varGamma_u \begin{bmatrix} a_{u, {\text T}}^* (\theta_{u, 1}, w_u)\\ \vdots \\ a_{u, {\text T}}^* (\theta_{u, L_u}, w_u) \end{bmatrix},
\end{align}
where $a_{u, {\text T}} (\theta_{u, l}, w_u) = e^{- j \frac{2 \pi}{\lambda} w_u \cos \theta_{u, l}}$. It can be proven that with one antenna at each FAS, every user should transmit at the maximum power in the optimal case, i.e., $q_u = P_u, \forall u \in {\cal U}$.\footnote{This can be easily proven by doing eigen-decomposition to (\ref{capacity}) and using reductio ad absurdum. Due to space limitations, the details are omitted here.} Therefore, the sum capacity (\ref{capacity}) can be rewritten as
\begin{equation}\label{capacity2}
C(w_{\cal U}) = \log \left| \sum_{u=1}^U P_u \bm g_u (w_u) \bm g_u^H (w_u) + \bm I_M \right|,
\end{equation}
where $w_{\cal U} = \{w_1, \dots, w_U\}$, and we only need to optimize the antenna positions by solving
\begin{align}\label{problem2}
\max_{0 \leq w_u \leq W_u,\forall u \in {\cal U}} C(w_{\cal U}).
\end{align}
In the following two subsections, we respectively consider the single-user and multiuser cases.

\subsection{Single-User Case}
If there is only one user, we further drop the user index `$u$'. The channel (\ref{G_u3}) and capacity (\ref{problem2}) thus become
\begin{align}
\bm g (w) & = \sqrt{M} \bm A_{\text R} \bm \varGamma \begin{bmatrix} a_{\text T}^* (\theta_1, w)\\ \vdots \\ a_{\text T}^* (\theta_L, w) \end{bmatrix}, \label{g_w}\\
C (w) & = \log \left| P \bm g (w) \bm g^H (w) + \bm I_M \right|,\label{capacity_su}
\end{align}
and we consider problem
\begin{align}\label{problem3}
\max_{0 \leq w \leq W} C(w).
\end{align}
Before solving (\ref{problem3}), we first provide an approximation in the following theorem.

\begin{theorem}\label{lemma1}
In the single-user case where the user uses a single-antenna FAS, if $M$ is large enough, then for any $w \in [0, W]$, $C (w)$ approaches
\begin{equation}\label{C_0}
C_0 = \log \left( \sum_{l=1}^L P M |\gamma_l|^2 + 1 \right),
\end{equation}
where $\gamma_l$ is channel gain of the $l$-th path.
\end{theorem}

\itshape \textbf{Proof:}  \upshape
See Appendix~\ref{prove_lemma1}.
\hfill $\Box$

Now we solve (\ref{problem3}). Note that if there is only one path, i.e., $L=1$, the optimal solution of (\ref{problem3}) can be obtained directly from Lemma~\ref{lemma01}. In the following, we show that if $L=2$, the optimal solution can also be obtained in closed form, and in the general multi-path case, a near-optimal solution can be easily found by a one-dimensional search.

\subsubsection{Two Paths}
If $L = 2$, denote $\rho = \frac{2 \pi}{\lambda} ( \cos \theta_1 - \cos \theta_2 )$, 
\begin{align}\label{Psi}
\bm \varPsi = P M \bm \varGamma^H \bm A_{\text R}^H \bm A_{\text R} \bm \varGamma \triangleq \begin{bmatrix} \psi_1 & \psi_2 \\ \psi_2^* & \psi_3 \end{bmatrix},
\end{align}
$\psi_2^{\text {Re}} = {\text {Re}} (\psi_2)$, and $\psi_2^{\text {Im}} = {\text {Im}} (\psi_2)$. Then, according to (\ref{g_w}) and (\ref{capacity_su}), $C(w)$ can be reformulated as
\begin{align}\label{capacity_2path}
	& C (w) \nonumber\\
	= & \log \!\left| \!P M \bm A_{\text R} \bm \varGamma \!\begin{bmatrix}\! a_{\text T}^* (\theta_1, w)\\ a_{\text T}^* (\theta_2, w) \!\end{bmatrix} \!\!\left[ a_{\text T} (\theta_1, w), a_{\text T} (\theta_2, w) \right] \!\bm \varGamma^H \!\!\bm A_{\text R}^H \!+\! \bm I_M \!\right| \nonumber\\
	= & \log \left( \left[ a_{\text T} (\theta_1, w), a_{\text T} (\theta_2, w) \right] \begin{bmatrix} \psi_1 & \psi_2 \\ \psi_2^* & \psi_3 \end{bmatrix} \begin{bmatrix} a_{\text T}^* (\theta_1, w)\\ a_{\text T}^* (\theta_2, w) \end{bmatrix} + 1 \right) \nonumber\\
	= & \log \left( e^{j \rho w} \psi_2^* + e^{- j \rho w} \psi_2 + \psi_1 + \psi_3 + 1 \right) \nonumber\\
	= & \log \left( 2 \psi_2^{\text {Im}} \sin (\rho w) + 2 \psi_2^{\text {Re}} \cos (\rho w) + \psi_1 + \psi_3 + 1 \right), \!\!\!
\end{align}
where the second step follows from using (\ref{O1O2_1}) and (\ref{Psi}). Thus, problem (\ref{problem3}) is equivalent to
\begin{align}\label{problem4}
\max_{0 \leq w \leq W}\psi_2^{\text {Im}} \sin (\rho w) + \psi_2^{\text {Re}} \cos (\rho w).
\end{align}
Note that we assume $\theta_1 \neq \theta_2$, since otherwise the channel reduces to the one-path case. Then, $\rho \neq 0$. Denote $\mu = \arctan \frac{\psi_2^{\text {Re}}}{\psi_2^{\text {Im}}} \in [-\pi/2, \pi/2]$ and 
\begin{equation}\label{w_hat}
	w_0 = \left\{\!\!\!
	\begin{array}{ll}
		\frac{\pi/2 - \mu}{\rho}, ~{\text {if}}~ \psi_2^{\text {Im}} > 0 ~{\text {and}}~ \rho > 0, \vspace{0.3em}\\
		\frac{- \pi/2 - \mu}{\rho}, ~{\text {if}}~ \psi_2^{\text {Im}} > 0 ~{\text {and}}~ \rho < 0,\vspace{0.3em}\\
		\frac{3 \pi/2 - \mu}{\rho}, ~{\text {if}}~ \psi_2^{\text {Im}} < 0 ~{\text {and}}~ \rho > 0,\vspace{0.3em}\\
		\frac{- 3 \pi/2 - \mu}{\rho}, ~{\text {if}}~ \psi_2^{\text {Im}} < 0 ~{\text {and}}~ \rho < 0.
	\end{array} \right.
\end{equation}
In the following theorem, we provide the optimal solution of (\ref{problem4}) or (\ref{problem3}) for the two-path case.

\begin{theorem}\label{theorem1}
In the two-path case, if $\psi_2^{\text {Im}} = 0$, the optimal solution of (\ref{problem3}) is given by
\begin{equation}\label{opt_w1}
w^* = \left\{\!\!\!
\begin{array}{ll}
0, ~{\text {if}}~ \psi_2^{\text {Re}} \geq 0, \vspace{0.3em}\\
\pi/ |\rho|, ~{\text {if}}~ \psi_2^{\text {Re}} < 0 ~{\text {and}}~ \pi/ |\rho| \in [0, W],\vspace{0.3em}\\
W, ~{\text {otherwise}}.
\end{array} \right.
\end{equation}
Otherwise, if $\psi_2^{\text {Im}} \neq 0$, the optimal solution is found as
\begin{equation}\label{opt_w2}
w^* = \left\{\!\!\!
\begin{array}{ll}
w_0, ~{\text {if}}~ w_0 \in [0, W], \vspace{0.3em}\\
0, ~{\text {if}}~ w_0 \notin [0, W] ~{\text {and}}~ C(0) \geq C(W),\vspace{0.3em}\\
W, ~{\text {otherwise}},
\end{array} \right.
\end{equation}
where $w_0$ is defined in (\ref{w_hat}).
\end{theorem}

\itshape \textbf{Proof:}  \upshape
See Appendix~\ref{prove_theorem1}.
\hfill $\Box$

\subsubsection{Multiple Paths}
Unlike the single/two-path case, if $L > 2$, it is intractable to obtain the optimal solution of (\ref{problem3}) in closed form. Using (\ref{O1O2_1}) and (\ref{g_w}), we rewrite $C (w)$ in (\ref{capacity_su}) as
\begin{equation}\label{Cw}
C (w) \!=\! \log \!\left(\! \left[ a_{\text T} (\theta_1, w), \dots, a_{\text T} (\theta_L, w) \right] \bm \varPsi \!\begin{bmatrix} a_{\text T}^* (\theta_1, w)\\ \vdots \\ a_{\text T}^* (\theta_L, w) \end{bmatrix} \!+\! 1 \!\right)\!,
\end{equation}
where $\bm \varPsi = P M \bm \varGamma^H \bm A_{\text R}^H \bm A_{\text R} \bm \varGamma$. Then, for any $w$, the value of $C (w)$ can be easily computed. We evenly discretize the interval $[0, W]$ into $K + 1$ points and obtain set $\{ 0, \epsilon, 2 \epsilon, \dots, W \}$, where $\epsilon = W/K$. Then, a near-optimal solution of (\ref{problem3}) can be easily found by using the exhaustive search (ES) (or one-dimensional search) method, i.e.,\footnote{Here we say that the solution is near-optimal since the variable $w$ is continuous and it is impossible to really perform ES. Therefore, we discretize the interval and carry out one-dimensional search, whose performance depends on the size of the search step $\epsilon$.}
\begin{equation}\label{find_w}
w^* = \arg \max_{w \in \{ 0, \epsilon, \dots, W \}} C (w).
\end{equation}

\subsection{Multiuser Case}
Now we consider the multiuser case and provide two schemes to solve problem (\ref{problem2}).

\subsubsection{Alternative Optimization}

Since problem (\ref{problem2}) is non-convex and intractable, we iteratively optimize $w_u, \forall u \in {\cal U}$.
For any $u \in {\cal U}$, denote ${\cal U}' = {\cal U} \setminus \{u\}$ and
\begin{align}
	\bm \varUpsilon_u & = \sum_{u' \in {\cal U}'} P_{u'} \bm g_{u'} (w_{u'}) \bm g_{u'}^H (w_{u'}) + \bm I_M, \label{omega_u} \\
	\bm \varPsi_u & = P_u M \bm \varGamma_u^H \bm A_{u, {\text R}}^H \bm \varUpsilon_u^{-1} \bm A_{u, {\text R}} \bm \varGamma_u. \label{psi_u}
\end{align}
Then, if $w_{u'}, \forall u' \in {\cal U}'$ are given, using (\ref{O1O2_1}) and (\ref{G_u3}), the sum capacity in (\ref{capacity2}) can be rewritten as
\begin{align}\label{ob_trans}
	C(w_u) & = \log \left| P_u \bm g_u (w_u) \bm g_u^H (w_u) \bm \varUpsilon_u^{-1} + \bm I_M \right| + \log \left| \bm \varUpsilon_u \right| \nonumber\\
	& = \log \Bigg( \left[ a_{u, {\text T}} (\theta_{u, 1}, w_u), \dots, a_{u, {\text T}} (\theta_{u, L_u}, w_u) \right] \bm \varPsi_u  \nonumber\\
	& \times \begin{bmatrix} a_{u, {\text T}}^* (\theta_{u, 1}, w_u)\\ \vdots \\ a_{u, {\text T}}^* (\theta_{u, L_u}, w_u) \end{bmatrix} + 1 \Bigg) + \log \left| \bm \varUpsilon_u \right|.
\end{align}
If $L_u = 1$, we know from Lemma~\ref{lemma0} that changing the antenna position of user~$u$ does not bring any capacity gain. As such, we simply let $w_u = 0$. If $L_u = 2$, then (\ref{ob_trans}) can be similarly expressed as (\ref{capacity_2path}). The optimal antenna position can therefore be obtained based on Theorem~\ref{theorem1}. If $L_u > 3$, we evenly discretize the interval $[0, W_u]$ into $K_u + 1$ points, obtain set $\{ 0, \epsilon_u, 2 \epsilon_u, \dots, W_u \}$, where $\epsilon_u = W_u/K_u$, and then find a near-optimal solution from this set by the ES method, i.e.,
\begin{equation}\label{find_wu0}
w_u^* = \arg \max_{w_u \in \{ 0, \epsilon_u, \dots, W_u \}} C (w_u).
\end{equation}
The main steps are summarized in the following Algorithm~\ref{algorithm1}.

\begin{algorithm}[h]
	\begin{algorithmic}[1]
		\caption{Alternative optimization for solving (\ref{problem2})}
		\State Initialize $w_{\cal U} = \{w_1, \cdots, w_U\}$.
		\Repeat
		\For{$u = 1:U$}
		\State Compute $\bm \varUpsilon_u$ and $\bm \varPsi_u$ based on (\ref{omega_u}) and (\ref{psi_u}).
		\If{$L_u = 1$}
		\State Let $w_u = 0$.
		\ElsIf{$L_u = 2$}
		\State Find the optimal $w_u$ based on Theorem~\ref{theorem1}.
		\Else
		\State Find the near-optimal $w_u$ based on (\ref{find_wu0}). 
		\EndIf
		\EndFor
		\Until{convergence}
		\label{algorithm1}
	\end{algorithmic}
\end{algorithm}

\subsubsection{Joint Optimization}
To avoid iterative and element-wise optimization of $w_u, \forall u \in {\cal U}$, we propose another algorithm below, which makes joint optimization of all variables possible. Most importantly, we will show in the next section that the technique can be modified and applied in solving the problem in the multi-antenna case.

Based on (\ref{G_u3}) and (\ref{capacity2}), (\ref{problem2}) can be rewritten as
\begin{align}\label{problem11}
\max_{0 \leq w_u \leq W_u, \forall u \in {\cal U}} \log \left| \sum_{u=1}^U\! P_u M \bm A_{u, {\text R}} \bm \varGamma_u \bm J_u (w_u) \bm \varGamma_u^H \!\bm A_{u, {\text R}}^H \!+\! \bm I_M \right|\!,
\end{align}
where 
\begin{align}
	& \bm J_u (w_u) = \nonumber\\
	& \begin{bmatrix}\! a_{u, {\text T}}^* (\theta_{u, 1}, w_u)\\ \vdots \\ a_{u, {\text T}}^* (\theta_{u, L_u}, w_u) \!\end{bmatrix} \!\!\left[ a_{u, {\text T}} (\theta_{u, 1}, w_u), \dots, a_{u, {\text T}} (\theta_{u, L_u}, w_u) \right]\!.\!\!\!\! \label{Jwu}
\end{align}
Noticing that $\bm J_u (w_u)$ is a rank-one PSD matrix with all $1$ diagonal elements, we replace $\bm J_u (w_u)$ in (\ref{problem11}) with another square matrix ${\hat{\bm J}}_u \in {\mathbb C}^{L_u \times L_u}$ and consider the following problem instead:
\begin{subequations}\label{problem12}
\begin{align}
	\mathop {\max }\limits_{{\hat{\bm J}}_{\cal U}} \quad & \log \left| \sum_{u=1}^U P_u M \bm A_{u, {\text R}} \bm \varGamma_u {\hat{\bm J}}_u \bm \varGamma_u^H \bm A_{u, {\text R}}^H + \bm I_M \right| \label{problem12_a}\\
	\text{s.t.} \quad\; & {\hat{\bm J}}_u \succeq \bm 0, ~\forall u \in {\cal U}, \label{problem12_b}\\
	& {\rm rank} ({\hat{\bm J}}_u) = 1, ~\forall u \in {\cal U}, \label{problem12_c}\\
	& {\rm diag} ({\hat{\bm J}}_u) = \bm 1, ~\forall u \in {\cal U},\label{problem12_d}
\end{align}
\end{subequations}
where ${\hat{\bm J}}_{\cal U} = \{{\hat{\bm J}}_1, \dots, {\hat{\bm J}}_U\}$. Obviously, any solution of (\ref{problem11}) corresponds to a feasible solution of (\ref{problem12}) (by simply letting ${\hat{\bm J}}_u = \bm J_u (w_u)$). However, the reverse does not hold, since all elements of $\bm J_u (w_u)$ are determined by the scalar variable $w_u$, while this is not a restriction for ${\hat{\bm J}}_u$. In the following, we first solve (\ref{problem12}), and then map its solution to that of (\ref{problem11}).

We replace the rank-one constraint (\ref{problem12_c}) with (\ref{problem13_c}) and arrive at the following equivalent problem:
\begin{subequations}\label{problem13}
\begin{align}
	\mathop {\max }\limits_{{\hat{\bm J}}_{\cal U}} \quad & \log \left| \sum_{u=1}^U P_u M \bm A_{u, {\text R}} \bm \varGamma_u {\hat{\bm J}}_u \bm \varGamma_u^H \bm A_{u, {\text R}}^H + \bm I_M \right| \label{problem13_a}\\
	\text{s.t.} \quad\; & {\hat{\bm J}}_u \succeq \bm 0, ~\forall u \in {\cal U}, \label{problem13_b}\\
	& \left\| {\hat{\bm J}}_u \right\|_* - \left\| {\hat{\bm J}}_u \right\|_2 \leq 0,~\forall u \in {\cal U}, \label{problem13_c}\\
	& {\rm diag} ({\hat{\bm J}}_u) = \bm 1, ~\forall u \in {\cal U},\label{problem13_d}
\end{align}
\end{subequations}
where $ \left\| \cdot \right\|_*$ and $\left\| \cdot \right\|_2$ are the nuclear and spectral norms of a matrix, which are, respectively, defined as the sum of all singular values of the matrix and the maximum singular value of the matrix. Since ${\hat{\bm J}}_u$ is a PSD matrix and is not a zero matrix (due to the constraint (\ref{problem13_d})), it is obvious that (\ref{problem13_c}) holds with equality if and only if ${\hat{\bm J}}_u$ is rank-one and is thus equivalent to (\ref{problem12_c}). Constraint (\ref{problem13_c}) takes on a difference of convex (DC) form and is thus non-convex. To overcome the non-convexity, we move (\ref{problem13_c}) to the objective function by the penalty-based method \cite[Ch. 17]{nocedal2006numerical} and therefore get
\begin{subequations}\label{problem14}
\begin{align}
	\mathop {\max }\limits_{{\hat{\bm J}}_{\cal U}} \quad & \log \left| \sum_{u=1}^U P_u M \bm A_{u, {\text R}} \bm \varGamma_u {\hat{\bm J}}_u \bm \varGamma_u^H \bm A_{u, {\text R}}^H + \bm I_M \right| \nonumber\\
	& - \sum_{u=1}^U \tau \left( \left\| {\hat{\bm J}}_u \right\|_* - \left\| {\hat{\bm J}}_u \right\|_2 \right) \label{problem14_a}\\
	\text{s.t.} \quad\; & {\hat{\bm J}}_u \succeq \bm 0, ~\forall u \in {\cal U}, \label{problem14_b}\\
	& {\rm diag} ({\hat{\bm J}}_u) = \bm 1, ~\forall u \in {\cal U}, \label{problem14_c}
\end{align}
\end{subequations}
where $\tau$ is the penalty factor penalizing the violation of (\ref{problem13_c}). Problem (\ref{problem14}) is a DC programming. The iterative majorization minimization (MM) based algorithm, which solves a sequence of convex problems by linearizing the non-convex objective function in each iteration, can thus be applied \cite{7547360}. Specifically, let ${\hat{\bm J}}_u^{(i)}$ denote the solution obtained in the previous iteration. By utilizing the first-order Taylor series approximation to linearize $\left\| {\hat{\bm J}}_u \right\|_2$, a lower bound can be obtained as  
\begin{equation}\label{J_star2_u}
	\left\| {\hat{\bm J}}_u \right\|_2 \geq \left\| {\hat{\bm J}}_u^{(i)} \right\|_2 + \bm v_u^H ({\hat{\bm J}}_u^{(i)}) ({\hat{\bm J}}_u - {\hat{\bm J}}_u^{(i)}) \bm v_u ({\hat{\bm J}}_u^{(i)}),
\end{equation}
where $\bm v_u ({\hat{\bm J}}_u^{(i)})$ is the eigenvector of ${\hat{\bm J}}_u^{(i)}$ associated with the maximum eigenvalue of ${\hat{\bm J}}_u^{(i)}$.
By replacing $\left\| {\hat{\bm J}}_u \right\|_2$ in (\ref{problem14_a}) with the lower bound, we solve (\ref{problem14}) by iteratively dealing with the following problem:
\begin{align}\label{problem15}
	\mathop {\max }\limits_{{\hat{\bm J}}_{\cal U}} \quad & \log \left| \sum_{u=1}^U P_u M \bm A_{u, {\text R}} \bm \varGamma_u {\hat{\bm J}}_u \bm \varGamma_u^H \bm A_{u, {\text R}}^H + \bm I_M \right| \nonumber\\
	& - \sum_{u=1}^U \tau \left[ \left\| {\hat{\bm J}}_u \right\|_* - \bm v_u^H ({\hat{\bm J}}_u^{(i)}) {\hat{\bm J}}_u \bm v_u ({\hat{\bm J}}_u^{(i)}) \right] \nonumber\\
	\text{s.t.} \quad\; & {\hat{\bm J}}_u \succeq \bm 0, ~\forall u \in {\cal U}, \nonumber\\
	& {\rm diag} ({\hat{\bm J}}_u) = \bm 1, ~\forall u \in {\cal U},
\end{align}
which is convex and can be optimally solved by CVX tool.

Once (\ref{problem12}) is solved, for each user~$u$, we look for the $w_u$ that minimizes the distance between $\bm J_u (w_u)$ and ${\hat{\bm J}}_u$, i.e.,
\begin{equation}\label{find_wu}
	w_u = \arg \min_{{\hat w}_u \in \{ 0, \epsilon_u, \dots, W_u \}} \left\| \bm J_u ({\hat w}_u) - {\hat{\bm J}}_u \right\|_F, ~\forall u \in {\cal U},
\end{equation}
where $\{ 0, \epsilon_u, \dots, W_u \}$ represents the quantization of the interval $[0, W_u]$. Then, we let $w_{\cal U}$ be the solution of (\ref{problem11}). The main steps are summarized in Algorithm~\ref{algorithm2}.

\begin{algorithm}[h]
	\begin{algorithmic}[1]
		\caption{Joint optimization for solving (\ref{problem11})}
		\State Initialize $w_{\cal U} = \{w_1, \cdots, w_U\}$ and $\tau$. Let $i = 0$.
		\State For any $u \in {\cal U}$, let ${\hat{\bm J}}_u^{(i)} = \bm J_u (w_u)$, which can be computed based on (\ref{Jwu}). 
		\Repeat
		\State Calculate $\bm v_u ({\hat{\bm J}}_u^{(i)}), \forall u \in {\cal U}$ and let $i = i + 1$.
		\State Solve problem (\ref{problem15}) by CVX and obtain ${\hat{\bm J}}_u^{(i)}, \forall u \in {\cal U}$.
		\Until{convergence}
		\State Let ${\hat{\bm J}}_u = {\hat{\bm J}}_u^{(i)}, \forall u \in {\cal U}$ be the solution of problem (\ref{problem12}).
		\State Obtain $w_{\cal U}$ from (\ref{find_wu}) and take it as the solution of (\ref{problem11}).
		\label{algorithm2}
	\end{algorithmic}
\end{algorithm}

\subsection{Convergence and Complexity Analysis}\label{convergence_complexity1}
In this subsection, we analyze the convergence and complexity of the proposed Algorithm~\ref{algorithm1} and Algorithm~\ref{algorithm2}.

\subsubsection{Convergence Analysis}
\label{convergence_analysis1}
As shown in Algorithm~\ref{algorithm1}, for each user~$u$, the problem of maximizing (\ref{ob_trans}) can either be optimally solved if $L_u \leq 2$ or near-optimally solved if $L_u > 2$. Hence, in each iteration, the objective function increases. Since the capacity is limited, the convergence of Algorithm~\ref{algorithm1} is guaranteed. The convergence of Algorithm~\ref{algorithm2} is determined by that of steps $3$ to $6$, which solve (\ref{problem14}) using the MM-based algorithm. From \cite{7547360} and \cite{7296696}, it is known that by successive convex approximation, the iterative steps will converge to a stationary point of (\ref{problem14}). Algorithm~\ref{algorithm2} thus converges.

\subsubsection{Complexity Analysis}
\label{complexity_analysis1}
To evaluate the complexity of the proposed algorithms, we count the total number of floating-point operations (FLOPs), where one FLOP represents a complex multiplication or summation. We express it as a polynomial function of the dimensions of the matrices involved, and simplify the expression by ignoring all terms except the leading (i.e., highest order or dominant) terms \cite{boyd2004convex, hunger2005floating}. It is worth mentioning that the given analysis only shows how the bounds on computational complexity are related to different problem dimensions. The actual load may vary depending on the structure simplifications and used numerical solvers.

For convenience, we assume $L_u = L$, $N_u = N$, and $K_u = K, \forall u \in {\cal U}$ when analyzing the complexity. One may also use $\max \{ L_u, \forall u \in {\cal U} \}$, $\max \{ N_u, \forall u \in {\cal U} \}$, and $\max \{ K_u, \forall u \in {\cal U} \}$ instead. The complexity of Algorithm~\ref{algorithm1} mainly lies in step~$10$, which looks for the near-optimal antenna position of user~$u$ by one-dimensional search. In this step, $C(w_u)$ in (\ref{ob_trans}) has to be computed $K+1$ times. Since the product of a $c_1 \times c_2$ dimensional matrix and a $c_2 \times c_3$ dimensional matrix costs ${\cal O}\left( c_1 c_2 c_3 \right)$ FLOPs, computing $C(w_u)$ requires a complexity of ${\cal O}\left( L^2 \right)$. Considering that step~$10$ has to be carried out $\kappa_1 U$ times in Algorithm~\ref{algorithm1}, where $\kappa_1$ is the number of outer iterations, the overall complexity of Algorithm~\ref{algorithm1} is ${\cal O}\left( \kappa_1 U (K+1) L^2 \right)$. The complexity of Algorithm~\ref{algorithm2} mainly lies in solving problem (\ref{problem15}). Since (\ref{problem15}) can be easily transformed to the general determinant maximization optimization problem \cite[(19)]{YANG2019100730} with $UL^2$ variables (i.e., all entries in ${\hat{\bm J}}_u , \forall u \in {\cal U}$), an $M$-dimensional matrix inside the determinant operation, and a $UL$-dimensional constraint space, using the results in \cite[(20)]{YANG2019100730} and \cite[Section~10]{Vandenberghe1998Determinant}, Algorithm~\ref{algorithm2} involves a total complexity of ${\cal O}\left( \kappa_2 \sqrt{U} U^2 L^3 \left( U^2 L^4 + M^2 \right)\right)$, where $\kappa_2$ is the number of iterations of Algorithm~\ref{algorithm2}.

\section{Multi-Antenna FAS}\label{multi_an_FAS}
Here, we consider the case where each FAS has multiple antennas. To ease our explanation, we start with the single-user case and then consider the general multiuser case.

\subsection{Single-User Case}
With only one user, we omit the user index `$u$' for brevity. Problem (\ref{problem1}) thus becomes
\begin{align}\label{problem16_Qw}
	\mathop {\max }\limits_{\bm Q, \bm w} \quad & \log | \bm G (\bm w) \bm Q \bm G^H (\bm w) + \bm I_M | \nonumber\\
	\text{s.t.} \quad\; & \bm Q \succeq \bm 0, ~ {\text {tr}} (\bm Q) \leq P, \nonumber\\
	& 0 \leq w_n \leq W, ~n \in {\cal N}, \nonumber\\
	& w_n \neq w_{n'}, ~\forall n, n' \in {\cal N},~ n \neq n'.
\end{align}
For any $(\bm Q, \bm w)$, according to the Reciprocity Lemma of a point-to-point MIMO channel \cite[Lemma~9.1]{el2011network}, we know that there always exists a PSD matrix $\bm F \in {\mathbb C}^{M \times M}$ such that ${\text {tr}} (\bm F) \leq P$ and
\begin{align}\label{capacity3}
	C(\bm Q, \bm w) & = \log | \bm G (\bm w) \bm Q \bm G^H (\bm w) + \bm I_M | \nonumber\\
	& = \log | \bm G^H (\bm w) \bm F \bm G (\bm w) + \bm I_N |.
\end{align}
Then, we consider the following problem instead:
\begin{align}\label{problem16}
	\mathop {\max }\limits_{\bm F, \bm w} \quad & \log | \bm G^H (\bm w) \bm F \bm G (\bm w) + \bm I_N | \nonumber\\
	\text{s.t.} \quad\; & \bm F \succeq \bm 0, ~ {\text {tr}} (\bm F) \leq P, \nonumber\\
	& 0 \leq w_n \leq W, ~n \in {\cal N}, \nonumber\\
	& w_n \neq w_{n'}, ~\forall n, n' \in {\cal N},~ n \neq n'.
\end{align}
In the following Theorem~\ref{theorem2} we show that the problems (\ref{problem16_Qw}) and (\ref{problem16}) are equivalent in the sense that their optimal objective function values are equal. To facilitate the statement below, we define two problems
\begin{align}
	\mathop {\max }\limits_{\bm Q} \quad & \log | \bm G (\bm w) \bm Q \bm G^H (\bm w) + \bm I_M | \nonumber\\
	\text{s.t.} \quad\; & \bm Q \succeq \bm 0, ~ {\text {tr}} (\bm Q) \leq P,\label{problem16_Qw1}\\
	\mathop {\max }\limits_{\bm F} \quad & \log | \bm G^H (\bm w) \bm F \bm G (\bm w) + \bm I_N | \nonumber\\
	\text{s.t.} \quad\; & \bm F \succeq \bm 0, ~ {\text {tr}} (\bm F) \leq P,\label{problem16_1}
\end{align}
which can be seen as the subproblems resulting from (\ref{problem16_Qw}) and (\ref{problem16}) with given $\bm w$.

\begin{theorem}\label{theorem2}
If $(\bm Q^*, \bm w^*)$ and $(\bm F^\divideontimes, \bm w^\divideontimes)$ are respectively the optimal solutions of (\ref{problem16_Qw}) and (\ref{problem16}), the following statements are true:
\begin{enumerate}
\item The optimal objective function values of (\ref{problem16_Qw}) and (\ref{problem16}) are equal, i.e.,
\begin{align}\label{eq1}
& \log | \bm G (\bm w^*) \bm Q^* \bm G^H (\bm w^*) + \bm I_M | \nonumber\\
= & \log | \bm G^H (\bm w^\divideontimes) \bm F^\divideontimes \bm G (\bm w^\divideontimes) + \bm I_N |.
\end{align}
\item Given $\bm w^*$, letting $\bm F^*$ be the optimal solution of (\ref{problem16_1}), then, $(\bm F^*, \bm w^*)$ is the optimal solution of (\ref{problem16}) and 
\begin{align}\label{eq2}
& \log | \bm G^H (\bm w^*) \bm F^* \bm G (\bm w^*) + \bm I_N | \nonumber\\
= & \log | \bm G^H (\bm w^\divideontimes) \bm F^\divideontimes \bm G (\bm w^\divideontimes) + \bm I_N |.
\end{align}
\item Given $\bm w^\divideontimes$, letting $\bm Q^\divideontimes$ be the optimal solution of (\ref{problem16_Qw1}), then, $(\bm Q^\divideontimes, \bm w^\divideontimes)$ is the optimal solution of (\ref{problem16_Qw}) and 
\begin{align}\label{eq3}
& \log | \bm G (\bm w^*) \bm Q^* \bm G^H (\bm w^*) + \bm I_M | \nonumber\\
= & \log | \bm G (\bm w^\divideontimes) \bm Q^\divideontimes \bm G^H (\bm w^\divideontimes) + \bm I_M |.
\end{align}
\end{enumerate}
\end{theorem}

\itshape \textbf{Proof:}  \upshape
See Appendix~\ref{prove_theorem2}.
\hfill $\Box$

Theorem~\ref{theorem2} indicates that we can first solve (\ref{problem16}) and then obtain a solution for the original problem (\ref{problem16_Qw}). Denote 
\begin{align}
	\bm \varPhi & = M N \bm \varGamma^H \bm A_{\text R}^H \bm F \bm A_{\text R} \bm \varGamma, \label{Phi}\\
	\bm J_n (w_n) & = \begin{bmatrix}\! a_{\text T}^* (\theta_1, w_n)\\ \vdots \\ a_{\text T}^* (\theta_L, w_n) \!\end{bmatrix}\! \left[ a_{\text T} (\theta_1, w_n), \dots, a_{\text T} (\theta_L, w_n) \right]\!. \label{Jn_wn}\!\!\!
\end{align}
Based on the definition of $\bm G (\bm w)$, the objective function of (\ref{problem16}) can be rewritten as
\begin{align}\label{capacity4}
	& \log | \bm G^H (\bm w) \bm F \bm G (\bm w) + \bm I_N | \nonumber\\
	= & \log | M N \bm A_{\text T} (\bm w) \bm \varGamma^H \bm A_{\text R}^H \bm F \bm A_{\text R} \bm \varGamma \bm A_{\text T}^H (\bm w) + \bm I_N | \nonumber\\
	\overset{\text {(a)}}{=} & \log | \bm A_{\text T}^H (\bm w) \bm A_{\text T} (\bm w) \bm \varPhi + \bm I_L | \nonumber\\
	= & \log \left| \sum_{n = 1}^N \bm J_n (w_n) \bm \varPhi + \bm I_L \right|,
\end{align}
where (a) follows from using (\ref{Phi}) and (\ref{O1O2_1}). Note that $\bm J_n (w_n)$ is a rank-one PSD matrix and all of its diagonal elements are $1$. Based on this observation, we make a relaxation to (\ref{problem16}).

In particular, we define another rank-one PSD matrix ${\hat{\bm J}}_n \in {\mathbb C}^{L \times L}$, whose  diagonal elements are all~$1 / N$.
With these properties, ${\hat{\bm J}}_n$ can be expressed as
\begin{equation}\label{Jb}
	{\hat{\bm J}}_n = \bm b_n \bm b_n^H,
\end{equation}
where $\bm b_n \in {\mathbb C}^{L \times 1}$ and all its elements have module $1 / N$. We further define
\begin{align}
	& {\hat {\bm A}}_{\text T} ({\hat{\bm J}}_{\cal N}) = \begin{bmatrix} \bm b_1^H \\ \vdots \\ \bm b_N^H \end{bmatrix}, \nonumber\\
	& {\hat {\bm G}} ({\hat{\bm J}}_{\cal N}) = \sqrt{M N} \bm A_{\text R} \bm \varGamma {\hat {\bm A}}_{\text T}^H ({\hat{\bm J}}_{\cal N}),
\end{align}
where ${\hat{\bm J}}_{\cal N} = \{{\hat{\bm J}}_1, \dots, {\hat{\bm J}}_N\}$. Here ${\hat {\bm A}}_{\text T} ({\hat{\bm J}}_{\cal N})$ and ${\hat {\bm G}} ({\hat{\bm J}}_{\cal N})$ can be seen as the approximations $\bm A_{\text T} (\bm w)$ and $\bm G (\bm w)$. Then, instead of (\ref{problem16}), we consider the following relaxed problem:
\begin{align}\label{problem16_3}
	\mathop {\max }\limits_{\bm F, {\hat{\bm J}}_{\cal N}} \quad & \log | {\hat {\bm G}}^H ({\hat{\bm J}}_{\cal N}) \bm F {\hat {\bm G}} ({\hat{\bm J}}_{\cal N}) + \bm I_N | \nonumber\\
	\text{s.t.} \quad\; & \bm F \succeq \bm 0, ~ {\text {tr}} (\bm F) \leq P, \nonumber\\
	&{\hat{\bm J}}_n \succeq \bm 0, ~\forall n \in {\cal N}, \nonumber\\
	& {\rm rank} ({\hat{\bm J}}_n) = 1, ~\forall n \in {\cal N}, \nonumber\\
	& {\rm diag} ({\hat{\bm J}}_n) = \bm 1 / N, ~\forall n \in {\cal N}.
\end{align}
Any solution to (\ref{problem16}) corresponds to a feasible solution of (\ref{problem16_3}), but the reverse does not hold. In the following we first solve (\ref{problem16_3}) and then map its solution to that of (\ref{problem16}). Then, a solution of (\ref{problem16_Qw}) can be obtained, given its equivalence to (\ref{problem16}).

Since (\ref{problem16_3}) is non-convex and intractable, we solve it by iteratively optimizing $\bm F$ and ${\hat{\bm J}}_{\cal N}$, i.e., dealing with 
\begin{align}\label{problem16_4}
	\mathop {\max }\limits_{\bm F} \quad & \log | {\hat {\bm G}}^H ({\hat{\bm J}}_{\cal N}) \bm F {\hat {\bm G}} ({\hat{\bm J}}_{\cal N}) + \bm I_N | \nonumber\\
	\text{s.t.} \quad\; & \bm F \succeq \bm 0, ~ {\text {tr}} (\bm F) \leq P,
\end{align}
and 
\begin{align}\label{problem16_5}
	\mathop {\max }\limits_{{\hat{\bm J}}_{\cal N}} \quad & \log | {\hat {\bm G}}^H ({\hat{\bm J}}_{\cal N}) \bm F {\hat {\bm G}} ({\hat{\bm J}}_{\cal N}) + \bm I_N | \nonumber\\
	\text{s.t.} \quad\; & {\hat{\bm J}}_n \succeq \bm 0, ~\forall n \in {\cal N}, \nonumber\\
	& {\rm rank} ({\hat{\bm J}}_n) = 1, ~\forall n \in {\cal N}, \nonumber\\
	& {\rm diag} ({\hat{\bm J}}_n) = \bm 1 / N, ~\forall n \in {\cal N},
\end{align}
in an alternative manner. For given ${\hat{\bm J}}_{\cal N}$, (\ref{problem16_4}) is convex and can be optimally solved using the water-filling method. For given $\bm F$, similar to (\ref{capacity4}), the objective function of (\ref{problem16_5}) can be rewritten as
\begin{align}\label{capacity7}
	\log | {\hat {\bm G}}^H ({\hat{\bm J}}_{\cal N}) \bm F {\hat {\bm G}} ({\hat{\bm J}}_{\cal N}) + \bm I_N | = \log \left| \sum_{n = 1}^N {\hat{\bm J}}_n \bm \varPhi + \bm I_L \right|,
\end{align}
based on which (\ref{problem16_5}) can be equivalently transformed to
\begin{align}\label{problem18}
	\mathop {\max }\limits_{{\hat{\bm J}}_{\cal N}} \quad & \log \left| \sum_{n = 1}^N {\hat{\bm J}}_n \bm \varPhi + \bm I_L \right| \nonumber\\
	\text{s.t.} \quad\; & {\hat{\bm J}}_n \succeq \bm 0, ~\forall n \in {\cal N}, \nonumber\\
	& {\rm rank} ({\hat{\bm J}}_n) = 1, ~\forall n \in {\cal N}, \nonumber\\
	& {\rm diag} ({\hat{\bm J}}_n) = \bm 1 / N, ~\forall n \in {\cal N}.
\end{align}
Obviously, (\ref{problem16_5}) and (\ref{problem18}) are equivalent. Problem (\ref{problem18}) is non-convex due to the rank-one constraint. However, noticing that it takes on a similar form as (\ref{problem13}), it can be efficiently solved by iteratively dealing with the following problem:
\begin{align}\label{problem21}
	\mathop {\max }\limits_{{\hat{\bm J}}_{\cal N}} \; & \left| \sum_{n = 1}^N {\hat{\bm J}}_n \bm \varPhi \!+\! \bm I_L \right|  \!-\! \sum_{n=1}^N \tau \left[ \left\| {\hat{\bm J}}_n \right\|_* \!-\! \bm v_n^H ({\hat{\bm J}}_n^{(i)}) {\hat{\bm J}}_n \bm v_n ({\hat{\bm J}}_n^{(i)}) \right] \nonumber\\
	\text{s.t.} \;\; & {\hat{\bm J}}_n \succeq \bm 0, ~\forall n \in {\cal N}, \nonumber\\
	& {\rm diag} ({\hat{\bm J}}_n) = \bm 1 / N, ~\forall n \in {\cal N},
\end{align}
where $\tau$ is the penalty factor and $\bm v_n ({\hat{\bm J}}_n^{(i)})$ is the eigenvector corresponding to the maximum eigenvalue of ${\hat{\bm J}}_n^{(i)}$.

Once (\ref{problem16_3}) is solved, we map its solution $(\bm F, {\hat{\bm J}}_{\cal N})$ to the ``best'' feasible solution $(\bm F, \bm w)$ of (\ref{problem16}). That is, we optimize $w_1, \dots, w_N$ one by one, and for each antenna~$n$, we find the $w_n$ that minimizes the distance between $\bm J_n (w_n)$ and ${\hat{\bm J}}_n$ as 
\begin{equation}\label{find_wn}
w_n = \arg \min_{{\hat w}_n \in \{ 0, \epsilon, \dots, W \} \setminus {\cal X}} \left\| \bm J_n ({\hat w}_n) - {\hat{\bm J}}_n \right\|_F, 
\end{equation}
where $\{ 0, \epsilon, \dots, W \}$ is a quantization of the interval $[0, W]$, and ${\cal X}$ is the record of antenna positions that have already been determined. With the obtained $\bm w$, we update $\bm Q$ by solving (\ref{problem16_Qw1}) using the water-filling method, and then take $(\bm Q, \bm w)$ as the solution of (\ref{problem16_Qw}). The details are summarized in Algorithm~\ref{algorithm3}.

\begin{algorithm}[h]
	\begin{algorithmic}[1]
		\caption{Algorithm for solving (\ref{problem16_Qw})}
		\State Initialize $\bm w = [w_1, \cdots, w_N]^T$ and $\tau$. Let ${\cal X} = \phi$.
		\State Compute $\bm G (\bm w)$ and $\bm J_n (w_n), \forall n \in {\cal N}$.
		\State Let ${\hat {\bm G}} ({\hat{\bm J}}_{\cal N}) = \bm G (\bm w)$ and ${\hat{\bm J}}_n = \bm J_n (w_n), \forall n \in {\cal N}$.
		\Repeat
		\State Update $\bm F$ by solving (\ref{problem16_4}) via the water-filling method.
		\State Let $i = 0$ and ${\hat{\bm J}}_n^{(i)} = {\hat{\bm J}}_n, \forall n \in {\cal N}$. 
		\Repeat
		\State Calculate $\bm v_n ({\hat{\bm J}}_n^{(i)}), \forall n \in {\cal N}$ and let $i = i + 1$.
		\State Solve (\ref{problem21}) by CVX and obtain ${\hat{\bm J}}_n^{(i)}, \forall n \in {\cal N}$.
		\Until{convergence}
		\State Let ${\hat{\bm J}}_n = {\hat{\bm J}}_n^{(i)}, \forall n \in {\cal N}$ be the solution of (\ref{problem16_5}).
		\Until{convergence}
		\For{$n = 1:N$}
		\State Obtain $w_n$ from (\ref{find_wn}) and let ${\cal X} = {\cal X} \cup \{w_n\}$.
		\EndFor
		\State Compute $\bm G (\bm w)$, update $\bm Q$ by solving (\ref{problem16_Qw1}) via the water-filling method, and take $(\bm Q, \bm w)$ as the solution of (\ref{problem16_Qw}).
		\label{algorithm3}
	\end{algorithmic}
\end{algorithm}
\vspace{-3mm}

\subsection{Multiuser Case}
We solve (\ref{problem1}) by iteratively updating $(\bm Q_u, \bm w_u)$. Denote ${\cal U}' = {\cal U} \setminus \{u\}$,
\begin{align}\label{Upsilon}
	\bm \varUpsilon_u = \sum_{u' \in {\cal U}'} \bm G_{u'} (\bm w_{u'}) \bm Q_{u'} \bm G_{u'}^H (\bm w_{u'}) + \bm I_M,
\end{align}
the eigen-decomposition of its inverse by $\bm \varUpsilon_u^{-1} = \bm S_u \bm \varLambda_u  \bm S_u^H$, and
\begin{align}\label{G_bar}
	{\overline {\bm G}}_u (\bm w_u) & = \bm \varLambda_u^{\frac{1}{2}}  \bm S_u^H \bm G_u (\bm w_u) \nonumber\\
	& = \sqrt{M N_u} \bm \varLambda_u^{\frac{1}{2}}  \bm S_u^H \bm A_{u,{\text R}} \bm \varGamma_u \bm A_{u,{\text T}}^H (\bm w_u).
\end{align}
The sum capacity in (\ref{capacity}) can thus be rewritten as
\begin{align}\label{capacity5}
	& C(\bm Q_{\cal U}, \bm w_{\cal U}) \nonumber\\
	= & \log \left| \bm G_u (\bm w_u) \bm Q_u \bm G_u^H (\bm w_u) \bm \varUpsilon_u^{-1} + \bm I_M \right| + \log |\bm \varUpsilon_u| \nonumber\\
	= & \log \left| {\overline {\bm G}}_u (\bm w_u) \bm Q_u {\overline {\bm G}}_u^H (\bm w_u) + \bm I_M \right| + \log |\bm \varUpsilon_u|,
\end{align}
where the last step follows from using (\ref{O1O2_1}) and (\ref{G_bar}). Then, for given $\bm Q_{u'}, \bm w_{u'}, \forall u' \in {\cal U}\setminus \{u\}$, problem (\ref{problem1}) can be equivalently transformed to
\begin{align}\label{problem23}
	\mathop {\max }\limits_{\bm Q_u, \bm w_u} \quad & \log \left| {\overline {\bm G}}_u (\bm w_u) \bm Q_u {\overline {\bm G}}_u^H (\bm w_u) + \bm I_M \right| \nonumber\\
	\text{s.t.} \quad\;\; & \bm Q_u \succeq \bm 0, ~ {\text {tr}} (\bm Q_u) \leq P_u, \nonumber\\
	& 0 \leq w_{u, n} \leq W_u, ~ n \in {\cal N}_u, \nonumber\\
	& w_{u, n} \neq w_{u, n'}, ~ n, n' \in {\cal N}_u,~ n \neq n'.
\end{align}
Obviously, (\ref{problem23}) can be solved similarly as (\ref{problem16_Qw}) by Algorithm~\ref{algorithm3}. The details for solving (\ref{problem1}) are in Algorithm~\ref{algorithm4}.

\begin{algorithm}[h]
	\begin{algorithmic}[1]
		\caption{Algorithm for solving (\ref{problem1})}
		\State Initialize $\bm w_{\cal U}$ and $\tau$. Given $\bm w_{\cal U}$, solve (\ref{problem1}) by using the iterative water-filling method and obtain $\bm Q_{\cal U}$.
		\State Compute the sum capacity and denote it by $C_1$.
		\Repeat
		\For{$u = 1:U$}
		\State Solve (\ref{problem23}) by Algorithm~\ref{algorithm3} (without the first step).
		\State Compute the sum capacity and denote it by $C_2$.
		\If{$C_2 > C_1$}
		\State Update $(\bm Q_u, \bm w_u)$ and let $C_1 = C_2$.
		\EndIf
		\EndFor
		\Until{convergence}
		\label{algorithm4}
	\end{algorithmic}
\end{algorithm}
In Algorithm~\ref{algorithm3}, we first optimize ${\hat{\bm J}}_{\cal N}$ and then map it to $\bm w$ according to (\ref{find_wn}).
Note that this step may not always make the capacity increase.
To ensure that Algorithm~\ref{algorithm4} converges, at the end of each iteration (step~8 of Algorithm~\ref{algorithm4}), we update $(\bm Q_u, \bm w_u)$ only if the capacity can be improved.

	\subsection{Convergence and Complexity Analysis}
	\label{convergence_complexity2}
	In this subsection, we analyze the convergence and complexity of the proposed Algorithm~\ref{algorithm3} and Algorithm~\ref{algorithm4}.

	\subsubsection{Convergence Analysis}
	\label{convergence_analysis2}
	The convergence of Algorithm~\ref{algorithm3} is determined by that of steps $4$ to $12$, which solve problem (\ref{problem16_3}) by iteratively optimizing $\bm F$ and ${\hat{\bm J}}_{\cal N}$ in (\ref{problem16_4}) and (\ref{problem16_5}).
	Note that in each outer iteration, the optimal solution of problem (\ref{problem16_4}) is first obtained by using the water-filling method, and a suboptimal solution of (\ref{problem16_5}) is then obtained by using the MM-based method.
	Since the objective function of (\ref{problem16_3}) is obviously limited, the convergence is thus guaranteed.
	In Algorithm~\ref{algorithm4}, problem (\ref{problem1}) is solved by iteratively dealing with (\ref{problem23}) using Algorithm~\ref{algorithm3}.
	Note that we have proven the convergence of Algorithm~\ref{algorithm3}, which is determined by its $4$-th to $12$-th steps.
	Although the mapping operation in the subsequent steps does not affect the convergence of Algorithm~\ref{algorithm3}, it may make the capacity decrease when this algorithm is iteratively performed in Algorithm~\ref{algorithm4}.
	Therefore, as we have stated in the previous subsection, at the end of each iteration (step~8 of Algorithm~\ref{algorithm4}), we update $(\bm Q_u, \bm w_u)$ only if the capacity can be improved, and this guarantees the convergence of the algorithm.

	\subsubsection{Complexity Analysis}
	\label{complexity_analysis2}
	The complexity of Algorithm~\ref{algorithm3} mainly lies in solving (\ref{problem21}), which takes on a similar form as (\ref{problem15}).
	It can thus be similarly analyzed that Algorithm~\ref{algorithm3} involves an overall complexity of 
\begin{equation}
{\cal O}\left( \kappa_3 \kappa_4 \sqrt{N} N^2 L^3 \left( N^2 L^4 + L^2 \right)\right), 
\end{equation}
which can be simplified as ${\cal O}\left( \kappa_3 \kappa_4 \sqrt{N} N^4 L^7\right)$ by keeping the highest-order term only.
	Here $\kappa_3$ and $\kappa_4$ are respectively the numbers of outer and inner iterations of Algorithm~\ref{algorithm3}.
	Note that to solve (\ref{problem1}), Algorithm~\ref{algorithm3} is carried out $\kappa_5 U$ times in Algorithm~\ref{algorithm4}, where $\kappa_5$ is the number of outer iterations of Algorithm~\ref{algorithm4}. 
	Therefore, Algorithm~\ref{algorithm4} requires an overall computational complexity of ${\cal O}\left( \kappa_3 \kappa_4 \kappa_5 U \sqrt{N} N^4 L^7\right)$.


\section{Simulation Results}\label{simul}

\begin{figure}
	\centering
	\includegraphics[scale=0.45]{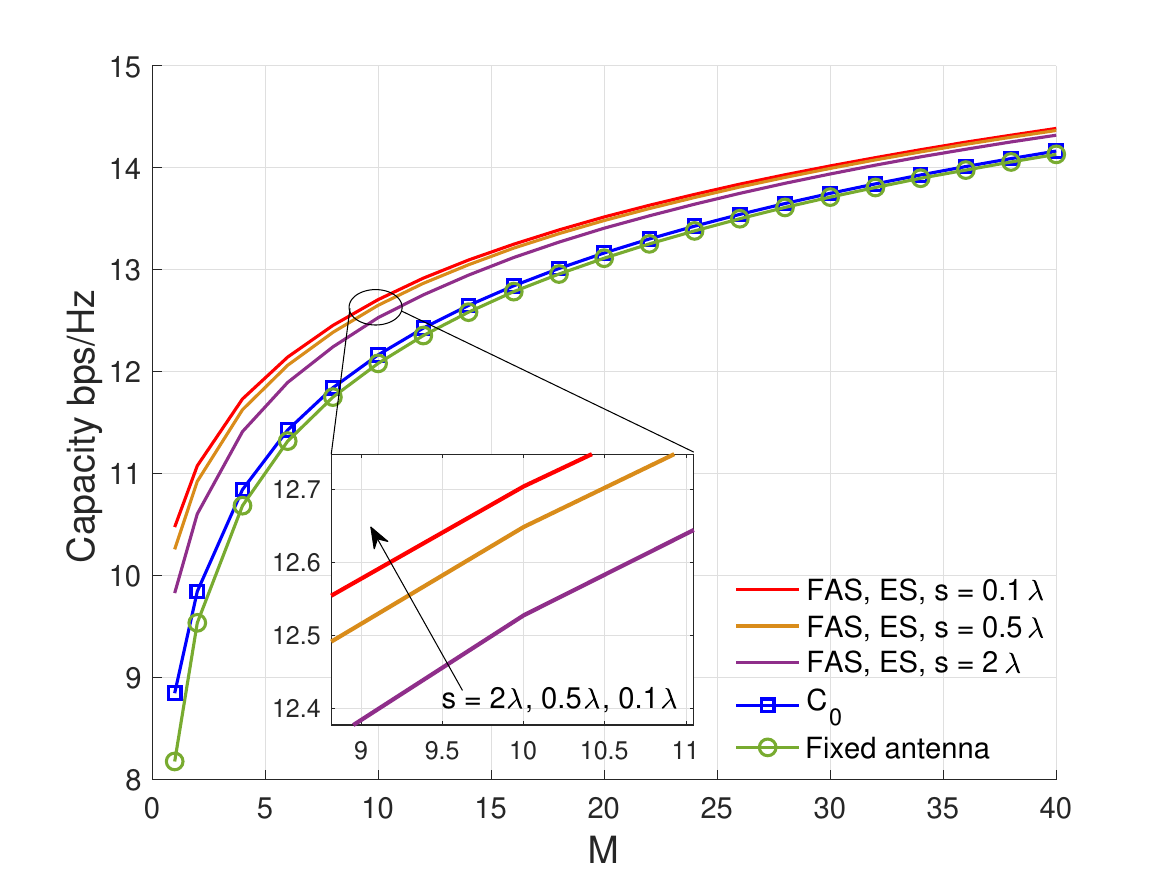}
	\vspace{-5mm}
	\caption{Single-antenna FAS \& Single-user: average capacity versus $M$ with $U=1$, $N=1$, ${\text {SNR}} =10$ dB, $W=10 \lambda$, and $L=5$.}
	\label{Capacity_VS_M_U1N1}
\end{figure}
Here, we assess the system performance by Monte Carlo simulations. For convenience, we assume $L_u = L$, $N_u = N$, $P_u = P$, $W_u = W$, and $K_u = K, \forall u \in {\cal U}$, which are respectively the number of propagation paths, the number of transmit antennas, maximum transmit power, FAS size, and quantization level of user~$u$. Since we consider normalized noise power in this paper, the signal-to-noise ratio (SNR) is $10 \log_{10} P$ dB. In the simulations, we set $K = 100$ and $\tau = 2$. The other parameters are specified in each figure. Besides the derived upper bound and approximations, we also compare the proposed algorithms with the following benchmarks:
\begin{itemize}
\item {\textbf {Fixed-antenna system}}: each user has $N$ fixed antennas with spacing $\frac{\lambda}{2}$.
Without loss of generality, we assume that the first antenna of each user locates at position~$0$.
Denote the capacity of this system by $C^{\text {fixed}}$, which can be obtained by the iterative water-filling method.
\item {\textbf {Exhaustive search (ES) method}}: we apply this method only in the single-antenna FAS case since in this scenario, all users should transmit at the maximum power, and we only need to optimize the antenna position of each user. We evenly discretize $[0, W]$ with $s$ being the step size, and then obtain the near-optimal solution by considering all possible combinations of $w_u, \forall u \in {\cal U}$.
\item {\textbf {Iterative water-filling and exhaustive search (IWF-ES) method}}: when a FAS has multiple antennas, we optimize $\bm Q_u$ and $\bm w_u$ iteratively by respectively using the water-filling and ES methods. When performing ES, we evenly discretize $[0, W]$ and then consider all possible combinations of $w_{u,n}, \forall n \in {\cal N}$.
Since the complexity of this step increases exponentially with $N$, we apply this method only in the single-user case (Fig.~\ref{Capacity_VS_M_U1N4}).
\item {\textbf {Simplified IWF-ES method}}: to reduce the computational complexity of the IWF-ES method, when optimizing $\bm w_u = [w_{u, 1}, \dots, w_{u, N}]^T$, instead of considering all possible combinations of $w_{u,n}, \forall n \in {\cal N}$, we optimize them one by one in an alternating manner.
\end{itemize}

\begin{figure}
	\centering
	\includegraphics[scale=0.45]{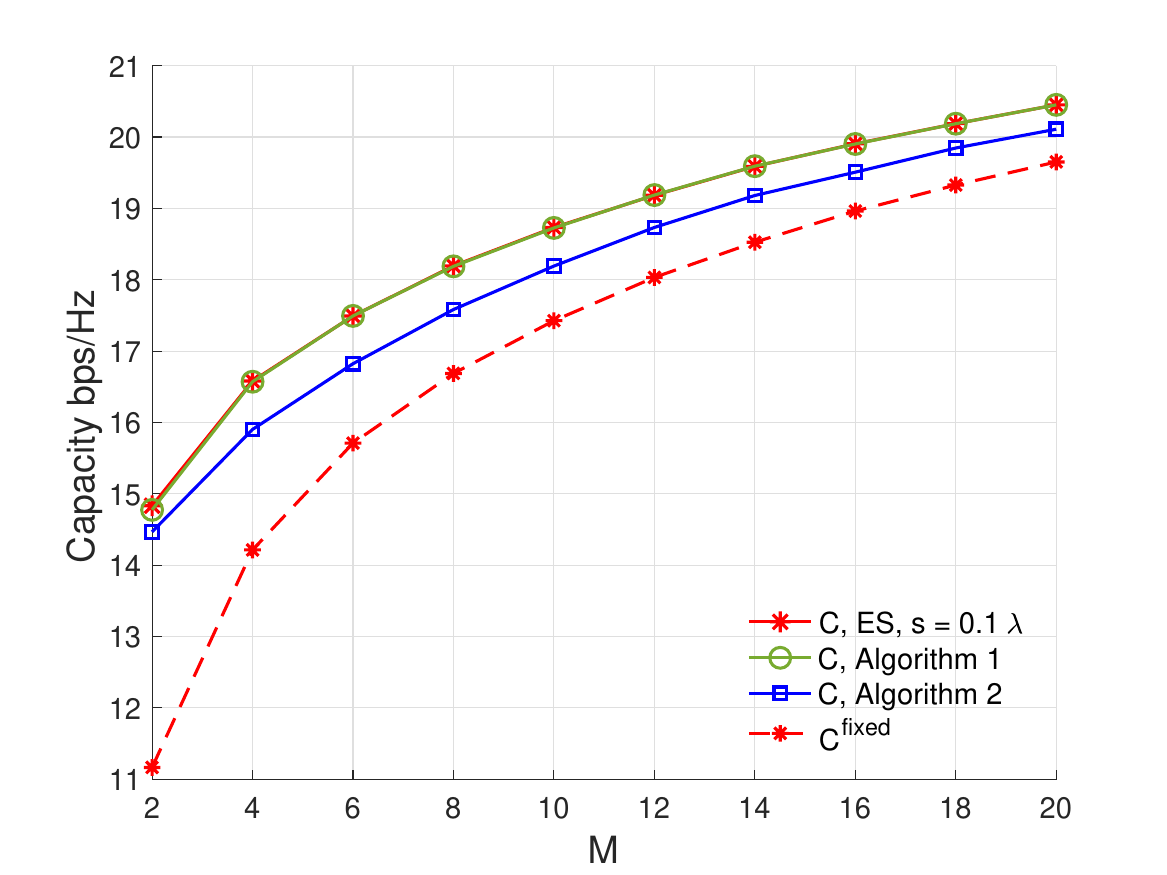}
	\vspace{-5mm}
	\caption{Single-antenna FAS \& Multiuser: average capacity versus $M$ with $U=2$, $N=1$, ${\text {SNR}}=10$ dB, $W=10 \lambda$, and $L=5$.}
	\label{Capacity_VS_M_U2N1}
\end{figure}
\begin{figure}
	\centering
	\includegraphics[scale=0.45]{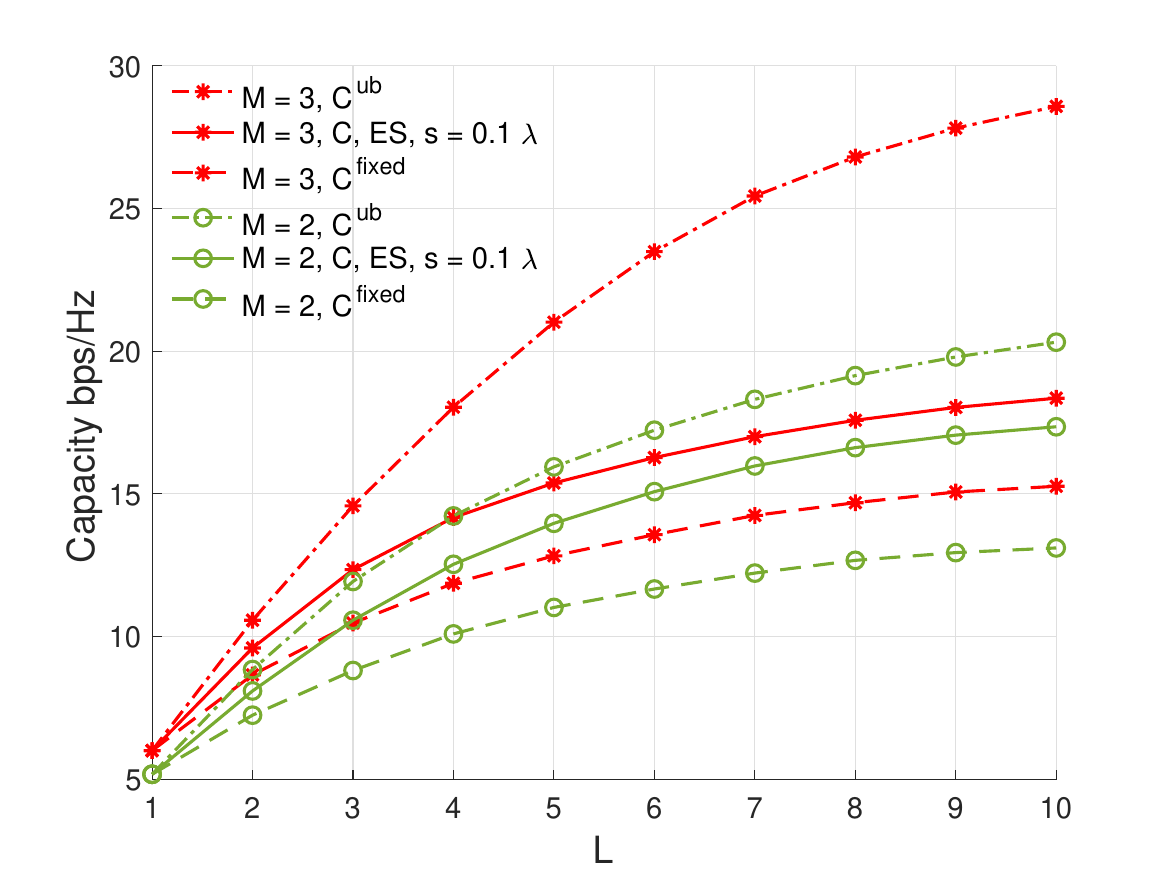}
	\vspace{-5mm}
	\caption{Single-antenna FAS \& Multiuser: average capacity versus $L$ with $U=2$, $N=1$, ${\text {SNR}}=10$ dB, and $W=10 \lambda$.}
	\label{Capacity_VS_L_ub}
\end{figure}

\subsection{Single-antenna FAS}

\subsubsection{Single-user Case}
Fig.~\ref{Capacity_VS_M_U1N1} compares the obtained $C$ with $C^{\text {fixed}}$ as well as $C_0$, which is derived in Theorem~\ref{lemma1}. It can be seen that when the search step size $s$ is $0.5 \lambda$, further reducing $s$ brings little capacity gain. When $M$ is small, we observe a large capacity gain by FAS. However, as $M$ increases, both $C$ and $C^{\text {fixed}}$ grow and approach $C_0$, and the effect of FAS becomes insignificant. This is consistent with Theorem~\ref{lemma1}.

\begin{figure}
\centering
\includegraphics[scale=0.45]{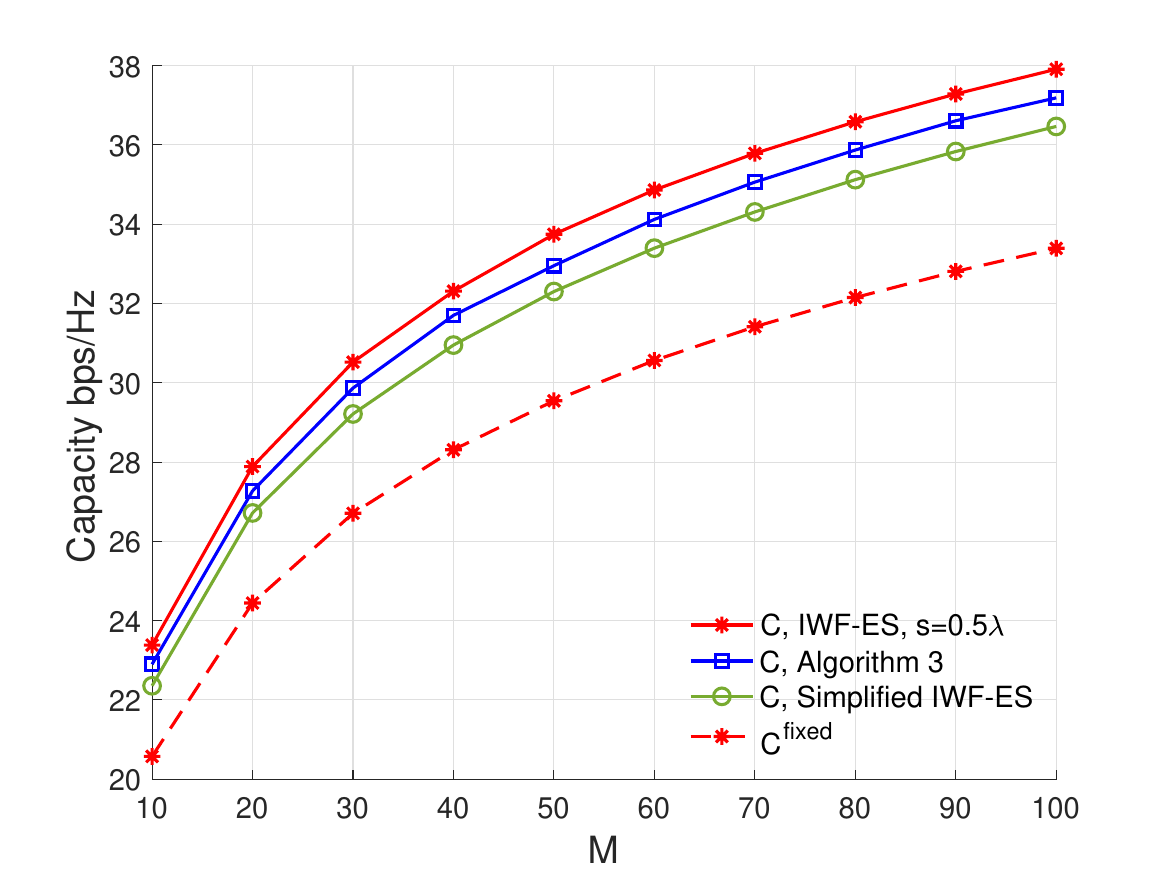}
\vspace{-5mm}
\caption{Multi-antenna FAS \& Single-user: average capacity versus $M$ with $U=1$, $N=4$, ${\text {SNR}}=10$ dB, $W=10 \lambda$, and $L=5$.}
\label{Capacity_VS_M_U1N4}
\end{figure}
\begin{figure}
	\centering
	\includegraphics[scale=0.45]{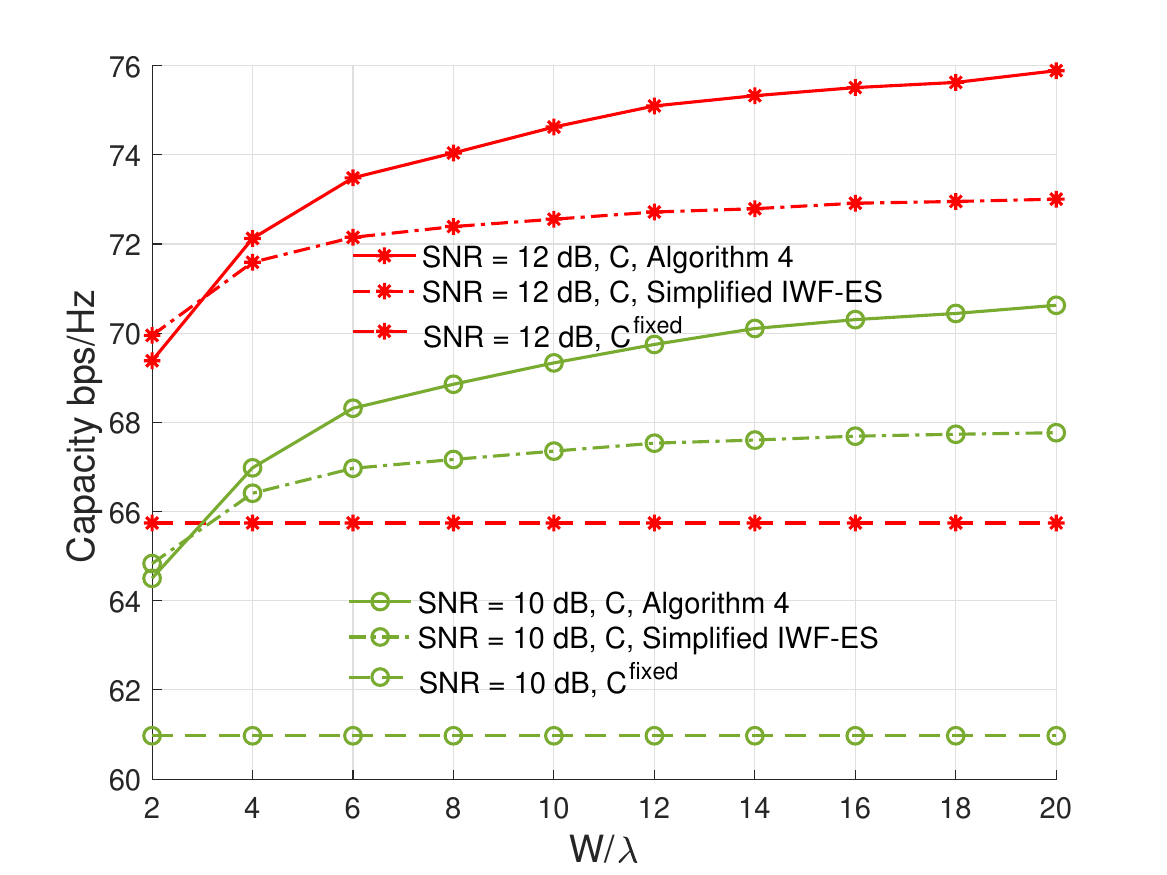}
	\vspace{-5mm}
	\caption{Multi-antenna FAS \& Multiuser: average capacity versus $W$ with $U=2$, $M=64$, $N=4$, and $L=5$.}
	\label{Capacity_VS_W_U2M64N4}
\end{figure}

\subsubsection{Multiuser Case}
Figs.~\ref{Capacity_VS_M_U2N1} and \ref{Capacity_VS_L_ub} consider the two-user case and respectively investigate the effect of $M$ and $L$. In Fig.~\ref{Capacity_VS_M_U2N1}, problem (\ref{problem2}) is solved by different schemes, i.e., the ES method, Algorithm~\ref{algorithm1}, and Algorithm~\ref{algorithm2}. It can be seen that the line obtained by Algorithm~\ref{algorithm1} almost coincides with that obtained by the ES method, and the line derived from Algorithm~\ref{algorithm2} also exhibits a remarkable proximity to the line obtained through the ES method, which shows the great performance of the proposed algorithms. Fig.~\ref{Capacity_VS_L_ub} mainly aims to evaluate the upper bound $C^{\text {ub}}$ derived in Theorem~\ref{theorem0}. For clarity, we only depict $C^{\text {fixed}}$, $C^{\text {ub}}$, and $C$ obtained by the ES method. It can be seen that when $L$ and $M$ are small, the gap between $C$ and $C^{\text {ub}}$ is small. Especially when $L = 1$, the gap disappears. This has been proven by Lemma~\ref{lemma01}. However, when $L$ or $M$ is large, the gap between $C$ and $C^{\text {ub}}$ is obvious, indicating that the upper bound is loose in this case. In the following, we consider relatively large $L$ and $M$. Therefore, we no longer depict $C^{\text {ub}}$, and we evaluate the performance of FAS and the proposed algorithms by other schemes.

\begin{figure}
\centering
\includegraphics[scale=0.45]{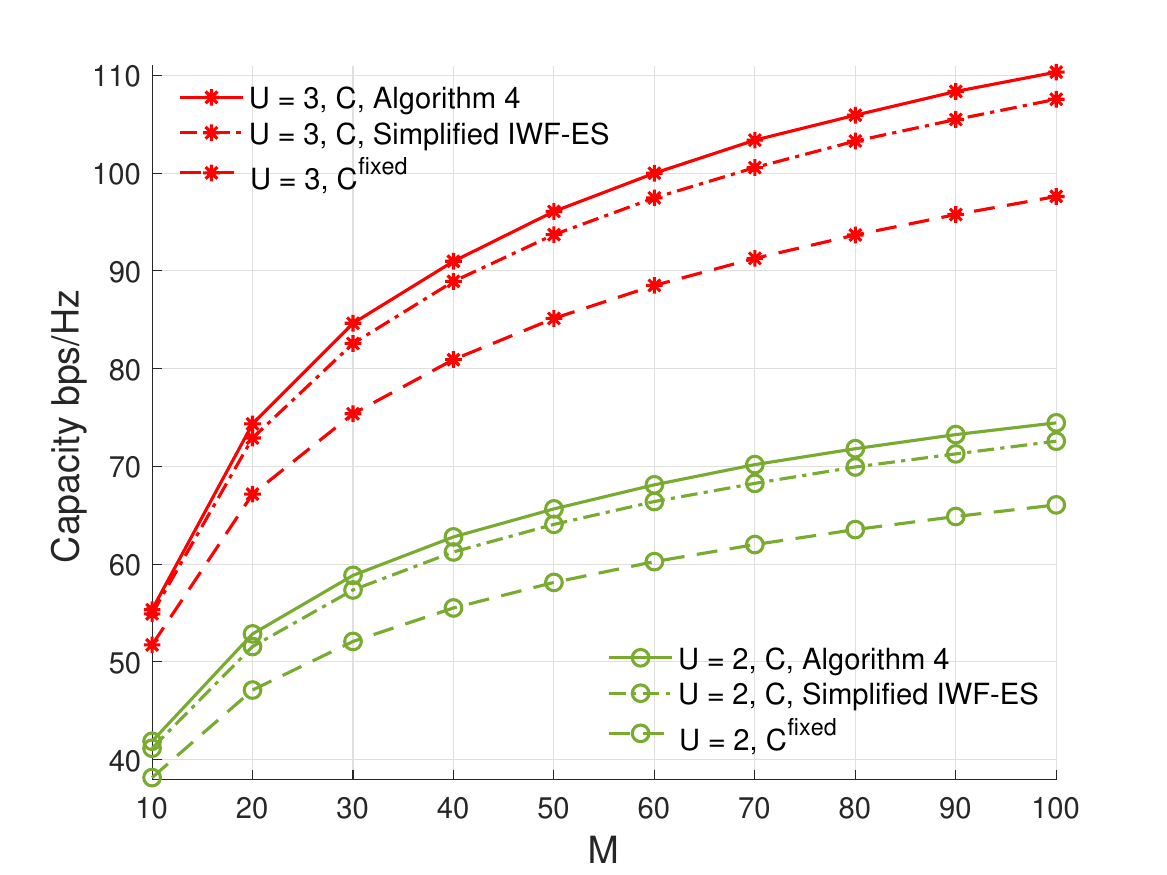}
\vspace{-5mm}
\caption{Multi-antenna FAS \& Multiuser: average capacity versus $M$ with $N=4$, ${\text {SNR}}=10$ dB, $W=10 \lambda$, and $L=5$.}
\label{Capacity_VS_M_U2N4}
\end{figure}
\begin{figure}
	\centering
	\includegraphics[scale=0.45]{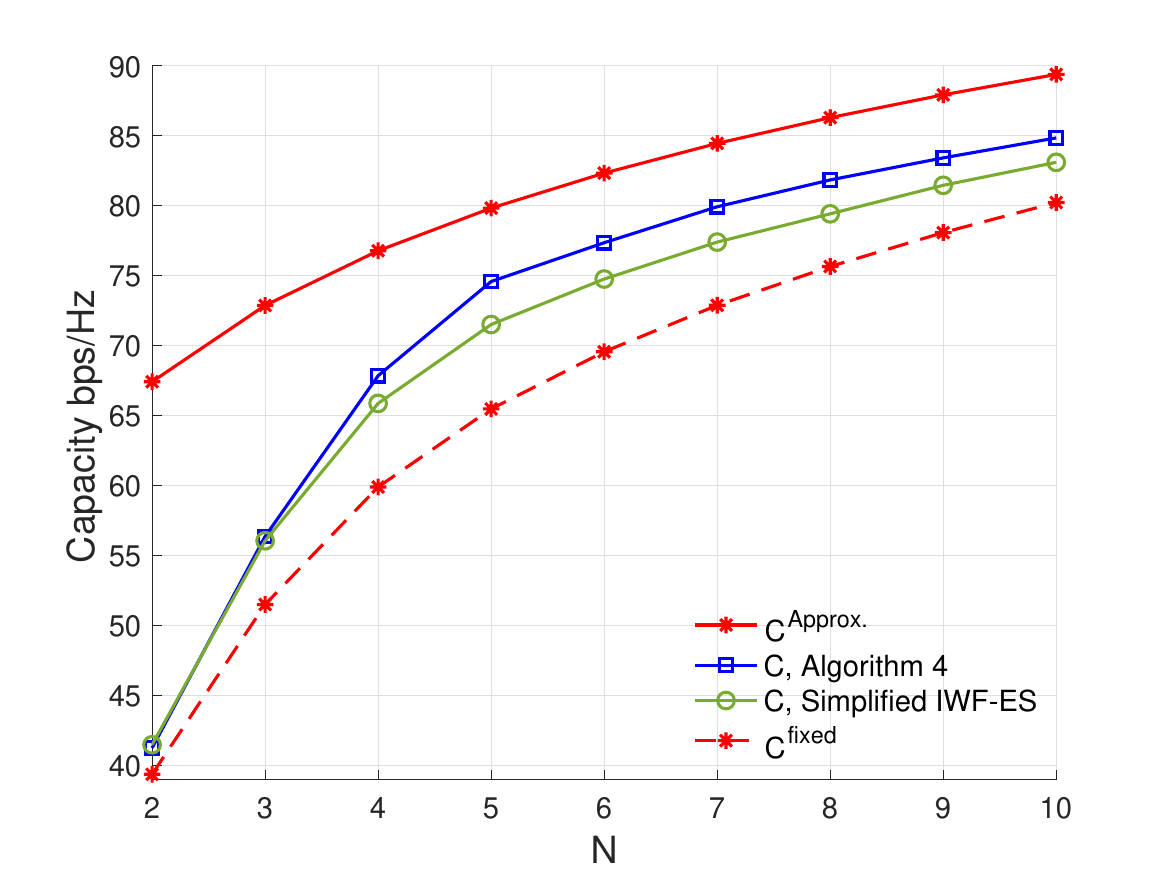}
	\vspace{-5mm}
	\caption{Multi-antenna FAS \& Multiuser: average capacity versus $N$ with $U=2$, $M=64$, ${\text {SNR}}=10$ dB, $W=10 \lambda$, and $L=5$.}
	\label{Capacity_VS_N_U2M64}
\end{figure}

\subsection{Multi-antenna FAS}
\subsubsection{Single-user Case}
Fig.~\ref{Capacity_VS_M_U1N4} considers solving (\ref{problem16_Qw}) by using Algorithm~\ref{algorithm3}, the IWF-ES method, and the simplified IWF-ES method. $C^{\text {fixed}}$ is also depicted as a benchmark. We see that the performance of Algorithm~\ref{algorithm3} is quite closed to that of IWF-ES, and better than the simplified IWF-ES method. 
In addition, unlike the single-antenna case, where the gap between $C$ and $C^{\text {fixed}}$ gets smaller and smaller as $M$ increases (see Fig.~\ref{Capacity_VS_M_U1N1}), in the multi-antenna case, the gap increases with $M$.
This can also be observed when there are multiple users (see Fig.~\ref{Capacity_VS_M_U2N4}), indicating that when the BS has a large number of antennas, the advantage of the multi-antenna FAS over the fixed-antenna system is more obvious than that of the single-antenna FAS.
This can be explained in an intuitive way as follows.
If a user uses a single-antenna FAS, adjusting the position of its antenna only changes the phase of the transmit steering element.
The channel gain brought by this is limited.
Thus, if $M$ is large, the capacity gain is mainly determined by the receive antennas.
In contrast, if a user uses a multi-antenna FAS, the system capacity benefits not only from the phase matching achieved by jointly adjusting the positions of the transmit antennas, but also from the transmit beamforming achieved by designing the covariance matrix.
Therefore, the capacity gain is determined by both the transmit and receive antennas.
Then, even if $M$ is large, using FAS can still bring significant capacity gain compared with the fixed antenna scheme.

\subsubsection{Multiuser Case}
Fig.~\ref{Capacity_VS_W_U2M64N4} investigates the effect of the FAS size $W$. As expected, the sum capacity of the system increases with $W$. Moreover, it is evident that with the increasing value of $W$, the capacity achieved by the simplified IWF-ES method converges rapidly, while that obtained through the proposed Algorithm~\ref{algorithm4} steadily increases. This discrepancy leads to a widening gap between the two methods.

In Figs.~\ref{Capacity_VS_M_U2N4} and \ref{Capacity_VS_N_U2M64}, the effects of $M$ and $N$ under different configurations are investigated. As anticipated, all results increase with $M$ and $N$. In addition, we can see that the gap between $C$ and $C^{\text {fixed}}$ increases with $M$ and $U$. Interestingly, as $N$ grows, the gap between $C$ and $C^{\text {fixed}}$ first increases and then decreases. This can be explained by Theorem~\ref{theorem3}, which proves that when $N$ is large, both $C$ and $C^{\text {fixed}}$ approach $C^{\rm approx}$.


\section{Conclusions}\label{conclusion}
This paper investigated the capacity of a FAS-assisted MAC. 
First, we derived upper bounds and approximations for the capacity, which not only provide valuable insights, but also serve as benchmarks for evaluating the performance of FAS. Then, we maximized the sum capacity by designing the transmit covariance matrices and antenna position vectors of the users. 
We showed that in some simple cases, the closed-form optimal solution exists, but in the general case, the problem is highly non-convex and intractable. Therefore, we proposed iterative algorithms to deal with the problem. The simulation results have validated the upper bounds and approximations on the sum capacity proposed in Section~\ref{section3} and Theorem~\ref{lemma1}. Moreover, the superior performance of the proposed algorithms over the considered benchmarks and the effectiveness of FAS in increasing the system capacity have also been verified. Note that unlike the traditional MIMO MAC system where all nodes use the fixed-position antennas, the problem in our case is much more complicated, and considering that FAS is a new topic, in this paper, we assume that all transmitters use FAS, but the receiver, i.e., the BS, uses the traditional fixed-position antennas. The investigation of the sum capacity maximization problem for the more general case where all nodes employ FAS remains a fascinating, yet challenging future work.

\appendices

\section{Condition for (\ref{O3O4}) to Hold with Equality}\label{equal_cond}
The inequality (\ref{O3O4}) is proven by \cite[Theorem~1]{coope1994matrix}. However, \cite{coope1994matrix} does not give the condition under which (\ref{O3O4}) holds with equality. The authors did not find the condition in anywhere else. Therefore, we prove the condition below.

First, we perform eigen-decomposition to $\bm O_3$ and $\bm O_4$, and obtain $\bm O_3 = \bm V_3 \bm D_3 \bm V_3^H$ and $\bm O_4 = \bm V_4 \bm D_4 \bm V_4^H$, where $\bm V_3$ and $\bm V_4$ are unitary matrices, and $\bm D_3$ and $\bm D_4$ are diagonal matrices consisting of the corresponding eigenvalues. Denote  $\bm \varDelta = \bm V_3 \bm D_3^{\frac{1}{2}} \bm V_3^H$ and $\bm \varOmega = \bm V_4 \bm D_4^{\frac{1}{2}} \bm V_4^H$. Then, $\bm O_3 = \bm \varDelta \bm \varDelta^H = \bm \varDelta^H \bm \varDelta$ and $\bm O_4 = \bm \varOmega \bm \varOmega^H = \bm \varOmega^H \bm \varOmega$. From the proof of \cite[Theorem~1]{coope1994matrix}, we directly have
\begin{align}\label{O3O4_2}
	{\text {tr}} (\bm O_3 \bm O_4) & = {\text {tr}} (\bm \varDelta \bm \varOmega^H \bm \varOmega \bm \varDelta^H) \nonumber\\
	& = \left\| \bm \varDelta \bm \varOmega^H \right\|_F^2 \nonumber\\
	& \leq \left\| \bm \varDelta \right\|_F^2 \left\| \bm \varOmega \right\|_F^2 \nonumber\\
	& = {\text {tr}} (\bm O_3) {\text {tr}} (\bm O_4).
\end{align}
The $(k,m)$-th element of the matrix $\bm \varDelta \bm \varOmega^H$ is $\sum_{r = 1}^N \delta_{k, r} \omega_{m, r}^*$, where $\delta_{k, r}$ and $\omega_{m, r}$ are respectively the $(k,r)$-th and $(m,r)$-th elements of $\bm \varDelta$ and $\bm \varOmega$. Then using the definition of Frobenius norm, the inequality in (\ref{O3O4_2}) can be written in detail as
\begin{align}\label{O3O4_3}
	\left\| \bm \varDelta \bm \varOmega^H \right\|_F^2 & = \sum_{k = 1}^N \sum_{m = 1}^N \left| \sum_{r = 1}^N \delta_{k, r} \omega_{m, r}^* \right|^2 \nonumber\\
	& \leq \sum_{k = 1}^N \sum_{m = 1}^N \left( \sum_{r = 1}^N |\delta_{k, r}|^2 \right) \left( \sum_{r = 1}^N |\omega_{m, r}|^2 \right) \nonumber\\
	& = \sum_{k = 1}^N \sum_{m = 1}^N \left( \sum_{r = 1}^N |\delta_{k, r}|^2 \right) \left( \sum_{t = 1}^N |\omega_{m, t}|^2 \right) \nonumber\\
	& = \left( \sum_{k = 1}^N \sum_{r = 1}^N |\delta_{k, r}|^2 \right) \left( \sum_{m = 1}^N \sum_{t = 1}^N |\omega_{m, t}|^2 \right) \nonumber\\
	& = \left\| \bm \varDelta \right\|_F^2 \left\| \bm \varOmega \right\|_F^2,
\end{align}
where the second step follows from the Cauchy-Schwarz inequality, i.e.,
\begin{align}\label{ineq1}
	\left| \sum_{r = 1}^N \delta_{k, r} \omega_{m, r}^* \right|^2 \leq & \left( \sum_{r = 1}^N |\delta_{k, r}|^2 \right) \left( \sum_{r = 1}^N |\omega_{m, r}|^2 \right), \forall k, m \in {\cal N},
\end{align}
and ${\cal N} = \{1, \dots, N\}$. For convenience, denote
\begin{align}
	\bm \delta_k & = [\delta_{k, 1}, \dots, \delta_{k, N}]^T, ~\forall k \in {\cal N},\\
	\bm \omega_m & = [\omega_{1, m}, \dots, \omega_{N, m}]^T, ~\forall m \in {\cal N},
\end{align}
which are respectively the transpose of the $k$-th row of $\bm \varDelta$ and the $m$-th row of $\bm \varOmega$. Thus, (\ref{ineq1}) can be equivalently rewritten as
\begin{equation}\label{ineq2}
	\left| \langle \bm \delta_k, \bm \omega_m \rangle \right|^2 \leq \langle \bm \delta_k, \bm \delta_k \rangle \langle \bm \omega_m, \bm \omega_m \rangle, \forall k, m \in {\cal N}.\!\!
\end{equation}
Note that for any given $k, m \in {\cal N}$, (\ref{ineq1}) or (\ref{ineq2}) holds with equality if and only if $\bm \delta_k$ and $\bm \omega_m$ are parallel.
Then, from (\ref{O3O4_2}) and (\ref{O3O4_3}), we know that (\ref{O3O4}) holds with equality if and only if any row of $\bm \varDelta$ is parallel to any row of $\bm \varOmega$.
Noticing that $\bm \varDelta$ (or $\bm \varOmega$) is a Hermitian matrix, all of its rows are thus parallel and all of its columns are also parallel.
Then, we know that $\bm \varDelta$ is rank-one and can be expressed as $\bm \varDelta = {\hat {\bm o}} {\hat {\bm o}}^H$, where ${\hat {\bm o}} \in {\mathbb C}^{N \times 1}$.
Since $\bm \varOmega$ is also Hermitian and any of its row is parallel to any row of $\bm \varDelta$, we know that it must be of the form $\bm \varOmega = \sqrt{\eta} {\hat {\bm o}} {\hat {\bm o}}^H$, where $\eta \in {\mathbb R}_+$ is a non-negative real constant.
Denote $\bm o = \sqrt{{\hat {\bm o}}^H {\hat {\bm o}}} {\hat {\bm o}}$.
Then, we have 
\begin{align}
	\bm O_3 & = \bm \varDelta \bm \varDelta^H = {\hat {\bm o}} {\hat {\bm o}}^H {\hat {\bm o}} {\hat {\bm o}}^H  = \bm o \bm o^H, \\
	\bm O_4 & = \bm \varOmega \bm \varOmega^H = \eta {\hat {\bm o}} {\hat {\bm o}}^H {\hat {\bm o}} {\hat {\bm o}}^H = \eta \bm o \bm o^H,
\end{align}
which is the necessary and sufficient condition for (\ref{O3O4}) to hold with equality. This completes the proof.

\section{Proof of Theorem~\ref{theorem0}}\label{prove_theorem0}
According to (\ref{G_u2}), the sum capacity of the system given in (\ref{capacity}) can be written in detail as
\begin{align}\label{capacity8}
	& C (\bm Q_{\cal U}, \bm w_{\cal U}) = \nonumber\\
	& \log \left| \sum_{u=1}^U M N_u \bm A_{u,{\text R}} \bm \varGamma_u \bm A_{u,{\text T}}^H (\bm w_u) \bm Q_u \bm A_{u,{\text T}} (\bm w_u) \bm \varGamma_u^H \!\bm A_{u,{\text R}}^H \!+\! \bm I_M \right|\!.
\end{align}
Given that $(\bm Q_{\cal U}, \bm w_{\cal U})$ is a feasible solution to (\ref{problem1}), let 
\begin{equation}\label{Q_tilde}
	{\tilde{\bm Q}}_u = \bm A_{u,{\text T}}^H (\bm w_u) \bm Q_u \bm A_{u,{\text T}} (\bm w_u), \forall u \in {\cal U}.
\end{equation}
Obviously, for any $u \in {\cal U}$, ${\tilde{\bm Q}}_u$ is PSD.
Moreover, it also satisfies the power constraint in (\ref{problem1_1}) since
\begin{align}\label{trace_ub}
	{\text {tr}} ({\tilde{\bm Q}}_u) & = {\text {tr}} (\bm A_{u,{\text T}}^H (\bm w_u) \bm Q_u \bm A_{u,{\text T}} (\bm w_u)) \nonumber\\
	& \overset{\text {(a)}}{=} {\text {tr}} (\bm A_{u,{\text T}} (\bm w_u) \bm A_{u,{\text T}}^H (\bm w_u) \bm Q_u) \nonumber\\
	& \overset{\text {(b)}}{\leq} {\text {tr}} (\bm A_{u,{\text T}} (\bm w_u) \bm A_{u,{\text T}}^H (\bm w_u)) {\text {tr}} (\bm Q_u) \nonumber\\
	& \overset{\text {(c)}}{=} {\text {tr}} (\bm A_{u,{\text T}}^H (\bm w_u) \bm A_{u,{\text T}} (\bm w_u)) {\text {tr}} (\bm Q_u) \nonumber\\
	& = L_u {\text {tr}} (\bm Q_u) \nonumber\\
	& \leq L_u P_u,
\end{align}
where (\ref{trace_ub}a), (\ref{trace_ub}b), and (\ref{trace_ub}c) follow from using (\ref{O1O2_2}), (\ref{O3O4}), and (\ref{O1O2_2}), respectively.
Therefore, ${\tilde{\bm Q}}_u$ is a feasible solution of (\ref{problem1_1}), indicating that from any feasible solution of (\ref{problem1}), by using (\ref{Q_tilde}), we could obtain a feasible solution for (\ref{problem1_1}). Note that the reverse does not necessarily hold true, i.e., from a feasible solution for (\ref{problem1_1}) we do not necessarily get a feasible solution for (\ref{problem1}). Denote the optimal solutions of (\ref{problem1}) and (\ref{problem1_1}) by $(\bm Q_{\cal U}^*, \bm w_{\cal U}^*)$ and ${\tilde{\bm Q}}_{\cal U}^*$, respectively. Then, we must have 
\begin{equation}
C (\bm Q_{\cal U}^*, \bm w_{\cal U}^*) \leq C^{\text {ub}} ({\tilde{\bm Q}}_{\cal U}^*),
\end{equation}
since otherwise, the solution of (\ref{problem1_1}) can be further improved based on (\ref{Q_tilde}). Theorem~\ref{theorem0} is thus proven.

\section{Proof of Lemma~\ref{lemma0}}\label{prove_lemma0}
Following similar steps in Appendix~\ref{prove_theorem0}, we can prove that the optimal objective function value of (\ref{problem1}) is upper bounded by that of the following problem:
\begin{align}\label{problem1_3}
\mathop {\max }\limits_{{\tilde{\bm Q}}_u, \bm Q_{{\cal U}'}, \bm w_{{\cal U}'}} \quad & \log \Bigg| \sum_{u' \in {\cal U}'} \bm G_{u'} (\bm w_{u'}) \bm Q_{u'} \bm G_{u'}^H (\bm w_{u'}) \nonumber\\
& \quad + M N_u \bm A_{u,{\text R}} \bm \varGamma_u {\tilde{\bm Q}}_u \bm \varGamma_u^H \bm A_{u,{\text R}}^H + \bm I_M \Bigg| \nonumber\\
\text{s.t.} \quad\quad\;\; & {\tilde{\bm Q}}_u \succeq \bm 0, ~ {\text {tr}} ({\tilde{\bm Q}}_u) \leq L_u P_u, \nonumber\\
& \bm Q_{u'} \succeq \bm 0, ~ {\text {tr}} (\bm Q_{u'}) \leq P_{u'}, ~\forall u' \in {\cal U}', \nonumber\\
& 0 \leq w_{u', n} \leq W_{u'}, ~\forall u' \in {\cal U}',~ n \in {\cal N}_{u'},
\end{align}
where ${\tilde{\bm Q}}_u \in {\mathbb C}^{L_u \times L_u}$. Different from Appendix~\ref{prove_theorem0} where the relaxation is made to all users, in (\ref{problem1_3}), it is made only to user~$u$. If $L_u = 1$, then $\bm A_{u,{\text R}}$ and $\bm \varGamma_u$ are reduced to $\bm a_{u,{\text R}} (\beta_{u, 1})$ and $\gamma_{u,1}$, respectively, and ${\tilde{\bm Q}}_u$ becomes a scalar variable within the interval $[0, P_u]$. It can be easily proven that in the optimal case, user~$u$ should transmit its signal at the maximum power, i.e., ${\tilde{\bm Q}}_u = P_u$, since otherwise, the objective function of (\ref{problem1_3}) can be further increased by increasing ${\tilde{\bm Q}}_u$. Then, (\ref{problem1_3}) becomes (\ref{problem1_2}), indicating that the optimal objective function value of (\ref{problem1}) is upper bounded by that of (\ref{problem1_2}). In the following we show that this upper bound is achievable and is thus tight. 

If $L_u = 1$, we know from (\ref{A_RT}) that
\begin{equation}\label{Aa}
	\bm A_{u,{\text T}} (\bm w_u) = \bm a_{u,{\text T}} (\theta_{u, 1}, \bm w_u).
\end{equation}
Let 
\begin{equation}\label{Q_u}
\bm Q_u = P_u \bm a_{u,{\text T}} (\theta_{u, 1}, \bm w_u) \bm a_{u,{\text T}}^H (\theta_{u, 1}, \bm w_u),
\end{equation}
which is obviously a PSD matrix and satisfies the maximum power constraint since 
\begin{align}
	{\text {tr}}(\bm Q_u) & = {\text {tr}}(P_u \bm a_{u,{\text T}} (\theta_{u, 1}, \bm w_u) \bm a_{u,{\text T}}^H (\theta_{u, 1}, \bm w_u)) \nonumber\\
	& = {\text {tr}}(P_u \bm a_{u,{\text T}}^H (\theta_{u, 1}, \bm w_u) \bm a_{u,{\text T}} (\theta_{u, 1}, \bm w_u)) \nonumber\\
	& = P_u.
\end{align}
Therefore, $\bm Q_u$ given by (\ref{Q_u}) satisfies the conditions on it in (\ref{problem1}).
Based on (\ref{Aa}) and (\ref{Q_u}) we know that for any $\bm w_u$,
\begin{align}
\bm A_{u,{\text T}}^H (\bm w_u) \bm Q_u \bm A_{u,{\text T}} (\bm w_u) & = P_u [ \bm a_{u,{\text T}}^H (\theta_{u, 1}, \bm w_u) \bm a_{u,{\text T}} (\theta_{u, 1}, \bm w_u) ]^2 \nonumber\\
& = P_u.
\end{align}
Then, the objective function of (\ref{problem1}) can be rewritten as
\begin{align}
	& C (\bm Q_{\cal U}, \bm w_{\cal U}) = \log \Bigg| \sum_{u' \in {\cal U}'} \bm G_{u'} (\bm w_{u'}) \bm Q_{u'} \bm G_{u'}^H (\bm w_{u'}) \nonumber\\
	& + M N_u \bm A_{u,{\text R}} \bm \varGamma_u \bm A_{u,{\text T}}^H (\bm w_u) \bm Q_u \bm A_{u,{\text T}} (\bm w_u) \bm \varGamma_u^H \bm A_{u,{\text R}}^H + \bm I_M \Bigg| \nonumber\\
	& = \log \Bigg| \sum_{u' \in {\cal U}'} \bm G_{u'} (\bm w_{u'}) \bm Q_{u'} \bm G_{u'}^H (\bm w_{u'}) \nonumber\\
	& + M N_u P_u |\gamma_{u,1}|^2 \bm a_{u,{\text R}} (\beta_{u, 1}) \bm a_{u,{\text R}}^H (\beta_{u, 1}) + \bm I_M \Bigg|,
\end{align}
which equals that of (\ref{problem1_2}). Thus, (\ref{problem1}) and (\ref{problem1_2}) are equivalent in the sense that their optimal function values are the same.

\section{Proof of Theorem~\ref{theorem3}}\label{prove_theorem3}
Let $(\bm Q_{\cal U}^*, \bm w_{\cal U}^*)$ and ${\hat {\bm Q}}_{\cal U}^*$ denote the optimal solutions of (\ref{problem1}) and (\ref{problem_approx}), respectively. In the following, we derive a lower bound and an upper bound to $C (\bm Q_{\cal U}^*, \bm w_{\cal U}^*)$, and show that if $N_u, \forall u \in {\cal U}$ are all large, both of them are approximate to $C^{\rm approx} ({\hat {\bm Q}}_{\cal U}^*)$. 
Therefore, we have 
\begin{equation}\label{appro1}
	C (\bm Q_{\cal U}^*, \bm w_{\cal U}^*) \approx C^{\rm approx} ({\hat {\bm Q}}_{\cal U}^*),
\end{equation}
which proves Theorem~\ref{theorem3}.

To obtain the lower bound, we assume that for each user~$u$, all of its $N_u$ antennas are equally spaced, and denote ${\bm w}_u' =  [w_{u, 1}', \dots, w_{u, N_u}']^T$, where $w_{u, n}' = \frac{W_u}{N_u - 1} (n - 1), n \in {\cal N}_u$.
Given ${\bm w}_u'$, problem (\ref{problem1}) reduces to 
\begin{align}\label{problem5}
	\mathop {\max }\limits_{\bm Q_{\cal U}} \quad & C(\bm Q_{\cal U}, \bm w_{\cal U}') \nonumber\\
	\text{s.t.} \quad\; & \bm Q_u \succeq \bm 0, ~ {\text {tr}} (\bm Q_u) \leq P_u, ~\forall u \in {\cal U},
\end{align}
where $\bm w_{\cal U}' = \{\bm w_1', \dots, \bm w_U'\}$.
Let $\bm Q_{\cal U}'$ denote the optimal solution of (\ref{problem5}).
Since ${\bm w}_u'$ is fixed in (\ref{problem5}), it is obvious that $C(\bm Q_{\cal U}', \bm w_{\cal U}')$ is a lower bound to $C (\bm Q_{\cal U}^*, \bm w_{\cal U}^*)$, i.e.,
\begin{equation}\label{lb}
	C(\bm Q_{\cal U}', \bm w_{\cal U}') \leq C (\bm Q_{\cal U}^*, \bm w_{\cal U}^*).
\end{equation}
In the following, we prove that if $N_u, \forall u \in {\cal U}$ are all large, for any feasible point $\bm Q_{\cal U}$ of (\ref{problem5}), we can always find a feasible point ${\hat {\bm Q}}_{\cal U}$ of (\ref{problem_approx}) such that $C(\bm Q_{\cal U}, \bm w_{\cal U}) \approx C^{\rm approx} ({\hat {\bm Q}}_{\cal U})$, and the reverse is also true.
Then, in the optimal case, we have
\begin{equation}\label{lb_approx}
	C(\bm Q_{\cal U}', \bm w_{\cal U}') \approx C^{\rm approx} ({\hat {\bm Q}}_{\cal U}^*).
\end{equation}

Since the elements in $\bm w_u'$ are equally spaced, when $N_u$ is large enough, the columns of $\bm A_{u,{\text T}} (\bm w_u')$ are approximately orthogonal to each other \cite[Lemma~1]{zhou2022channel}, i.e.,
\begin{equation}\label{orthogonal}
\bm a_{u, {\text T}}^H (\theta_{u, l}, \bm w_u') \bm a_{u, {\text T}} (\theta_{u, l'}, \bm w_u') \approx \left\{\!\!\!
\begin{array}{ll}
1, ~{\text {if}}~ l = l', \vspace{0.3em}\\
0, ~{\text {if}}~ l \neq l',
\end{array} \right.
\end{equation}
based on which we get
\begin{equation}\label{approx1}
\bm A_{u,{\text T}} (\bm w_u')^H \bm A_{u,{\text T}} (\bm w_u') \approx \bm I_{L_n}.
\end{equation}
We prove the approximation by separately discussing different users. Without loss of generality, we start from user~$1$. For convenience, denote
\begin{equation}\label{theta1}
\bm \varTheta_1 = \sum_{u' = 2}^U \bm G_{u'} (\bm w_{u'}') \bm Q_{u'} \bm G_{u'}^H (\bm w_{u'}') + \bm I_M,
\end{equation}
which is a PSD matrix.
The sum capacity of the system can thus be rewritten as
\begin{align}\label{capacity6}
	& C(\bm Q_{\cal U}, \bm w_{\cal U}') \nonumber\\
	= & \log \left| \bm G_1 (\bm w_1') \bm Q_1 \bm G_1^H (\bm w_1') + \bm \varTheta_1 \right| \nonumber\\
	= & \log | \bm \varTheta_1^{- \frac{1}{2}} \bm G_1 (\bm w_1') \bm Q_1 \bm G_1^H (\bm w_1') (\bm \varTheta_1^{- \frac{1}{2}})^H  + \bm I_M | + \log | \bm \varTheta_1 |,
\end{align}
where the last step follows from using (\ref{O1O2_0}) and (\ref{O1O2_1}). Note that the first log-determinant of (\ref{capacity6}) can be seen as the capacity of a point-to-point MIMO channel with channel $\bm \varTheta_1^{- \frac{1}{2}} \bm G_1 (\bm w_1')$ and covariance matrix $\bm Q_1$. According to the Reciprocity Lemma of a point-to-point MIMO channel \cite[Lemma~9.1]{el2011network}, we know that for any given $\bm Q_1$, there always exists a PSD matrix $\bm F_1 \in {\mathbb C}^{M \times M}$ such that ${\text {tr}} (\bm F_1) \leq P_1$ and (\ref{capacity6}) can be transformed and approximated as
\begin{align}\label{approx4}
	& C(\bm Q_{\cal U}, \bm w_{\cal U}') \nonumber\\
	= & \log | \bm G_1^H (\bm w_1') (\bm \varTheta_1^{- \frac{1}{2}})^H \bm F_1 \bm \varTheta_1^{- \frac{1}{2}} \bm G_1 (\bm w_1') + \bm I_{N_1} | + \log | \bm \varTheta_1 | \nonumber\\
	= & \log \!| M N_1 \bm A_{1,{\text T}} (\bm w_1') \bm \varGamma_1^H\! \bm A_{1,{\text R}}^H (\!\bm \varTheta_1^{\!- \frac{1}{2}}\!)^{\!H} \!\bm F_1 \bm \varTheta_1^{\!- \frac{1}{2}} \!\bm A_{1,{\text R}} \bm \varGamma_1 \bm A_{1,{\text T}}^H (\bm w_1') \nonumber\\
	& + \bm I_{N_1} | + \log | \bm \varTheta_1 | \nonumber\\
	\overset{\text {(a)}}{=} & \log \!| M N_1 \bm \varGamma_1^H\! \bm A_{1,{\text R}}^H (\!\bm \varTheta_1^{\!- \frac{1}{2}}\!)^{\!H} \!\bm F_1 \bm \varTheta_1^{\!- \frac{1}{2}} \!\bm A_{1,{\text R}} \bm \varGamma_1 \bm A_{1,{\text T}}^H (\bm w_1') \bm A_{1,{\text T}} (\bm w_1') \nonumber\\
	& + \bm I_{L_1} | + \log | \bm \varTheta_1 | \nonumber\\
	\overset{\text {(b)}}{\approx} & \log | M N_1 \bm \varGamma_1^H\! \bm A_{1,{\text R}}^H (\bm \varTheta_1^{\!- \frac{1}{2}})^H \!\bm F_1 \bm \varTheta_1^{\!- \frac{1}{2}} \!\bm A_{1,{\text R}} \bm \varGamma_1 \!+\! \bm I_{L_1} | \!+\! \log | \bm \varTheta_1 |.
\end{align}
where (a) and (b) respectively follow from using (\ref{O1O2_1}) and (\ref{approx1}). Again, based on the Reciprocity Lemma, we can always find a PSD matrix ${\hat {\bm Q}}_1 \in {\mathbb C}^{L_1 \times L_1}$ such that ${\text {tr}} ({\hat {\bm Q}}_1) \leq P_1$ and (\ref{approx4}) can be equivalently transformed to
\begin{align}\label{approx4_1}
	& \log | M N_1 \bm \varGamma_1^H\! \bm A_{1,{\text R}}^H (\bm \varTheta_1^{\!- \frac{1}{2}})^H \!\bm F_1 \bm \varTheta_1^{\!- \frac{1}{2}} \!\bm A_{1,{\text R}} \bm \varGamma_1 \!+\! \bm I_{L_1} | \!+\! \log | \bm \varTheta_1 | \nonumber\\
	= & \log | M N_1 \bm \varTheta_1^{\!- \frac{1}{2}} \!\bm A_{1,{\text R}} \bm \varGamma_1 \!{\hat {\bm Q}}_1 \bm \varGamma_1^H\! \bm A_{1,{\text R}}^H (\bm \varTheta_1^{\!- \frac{1}{2}})^H \!\!+\! \bm I_M | \!+\! \log | \bm \varTheta_1 | \nonumber\\
	= & \log | M N_1 \bm A_{1,{\text R}} \bm \varGamma_1 {\hat {\bm Q}}_1 \bm \varGamma_1^H \bm A_{1,{\text R}}^H + \bm \varTheta_1 |.
\end{align}
Combining (\ref{approx4}) and (\ref{approx4_1}), we know that if $N_1$ is large,
\begin{equation}
C(\bm Q_{\cal U}, \bm w_{\cal U}') \approx \log | M N_1 \bm A_{1,{\text R}} \bm \varGamma_1 {\hat {\bm Q}}_1 \bm \varGamma_1^H \bm A_{1,{\text R}}^H + \bm \varTheta_1 |.
\end{equation}

Now we consider user~$2$. Let
\begin{align}\label{theta2}
\bm \varTheta_2 & = M N_1 \bm A_{1,{\text R}} \bm \varGamma_1 {\hat {\bm Q}}_1 \bm \varGamma_1^H \bm A_{1,{\text R}}^H \nonumber\\
& + \sum_{u' = 3}^U \bm G_{u'} (\bm w_{u'}') \bm Q_{u'} \bm G_{u'}^H (\bm w_{u'}') + \bm I_M.
\end{align}
If $N_2$ is large, then we can prove by following similar steps that for any $\bm Q_2$, there exists a PSD matrix ${\hat {\bm Q}}_2 \in {\mathbb C}^{L_2 \times L_2}$ such that ${\text {tr}} ({\hat {\bm Q}}_2) \leq P_2$ and
\begin{align}\label{capacity9}
C(\bm Q_{\cal U}, \bm w_{\cal U}') & \approx \log \left| M N_2 \bm A_{2,{\text R}} \bm \varGamma_2 {\hat {\bm Q}}_2 \bm \varGamma_2^H \bm A_{2,{\text R}}^H + \bm \varTheta_2 \right| \nonumber\\
& = \log \left| \sum_{u = 1}^2 M N_u \bm A_{u,{\text R}} \bm \varGamma_u {\hat {\bm Q}}_u \bm \varGamma_u^H \bm A_{u,{\text R}}^H \right. \nonumber\\
& + \left. \sum_{u' = 3}^U \bm G_{u'} (\bm w_{u'}') \bm Q_{u'} \bm G_{u'}^H (\bm w_{u'}') + \bm I_M \right|.
\end{align}
By analogy, we know that if $N_u, \forall u \in {\cal U}$ are all large, for any feasible solution $\bm Q_{\cal U}$ of (\ref{problem5}), there always exists a feasible point ${\hat {\bm Q}}_{\cal U}$ of (\ref{problem_approx}) such that
\begin{align}\label{approx2}
C(\bm Q_{\cal U}, \bm w_{\cal U}') & \approx \log \left| \sum_{u = 1}^U M N_u \bm A_{u,{\text R}} \bm \varGamma_u {\hat {\bm Q}}_u \bm \varGamma_u^H \bm A_{u,{\text R}}^H + \bm I_M \right| \nonumber\\
& \triangleq C^{\rm approx} ({\hat {\bm Q}}_{\cal U}).
\end{align}
Following similar steps, it can be further proven that if $N_u, \forall u \in {\cal U}$ are all large, for any feasible point ${\hat {\bm Q}}_{\cal U}$ of problem (\ref{problem_approx}), there exist PSD matrices $\bm Q_u \in {\mathbb C}^{N_u \times N_u}, \forall u \in {\cal U}$ such that $\bm Q_{\cal U}$ is a feasible point of (\ref{problem5}), and the approximation in (\ref{approx2}) still holds.
Then, (\ref{lb_approx}) is true.

Next, we derive an upper bound to $C (\bm Q_{\cal U}^*, \bm w_{\cal U}^*)$, and show that if $N_u, \forall u \in {\cal U}$ are all large, this bound also approaches $C^{\rm approx} ({\hat {\bm Q}}_{\cal U})$. Once $(\bm Q_{\cal U}^*, \bm w_{\cal U}^*)$ is obtained, we know the optimal antenna positions of all users. Note that the antennas of each user may not be equally spaced. Then, (\ref{orthogonal}) cannot be directly used. To apply the above approximation technique, we can always construct a new system, where each user~$u$ has ${\hat N}_u \geq N_u$ equally spaced antennas and $N_u$ of them are located at the positions defined by $\bm w_{\cal U}^*$. All the other settings are the same as the system considered in this paper. 
Denote ${\bm w}_u'' =  [w_{u, 1}'', \dots, w_{u, {\hat N}_u}'']^T$, where $w_{u, n}'' = \frac{W_u}{{\hat N}_u - 1} (n - 1), n \in {\hat {\cal N}}_u$ and $N_u$ of them construct $\bm w_{\cal U}^*$, and consider the following problem:
\begin{align}\label{problem6}
	\mathop {\max }\limits_{{\overline {\bm Q}}_{\cal U}} \quad & {\overline C} ({\overline {\bm Q}}_{\cal U}, \bm w_{\cal U}'') \nonumber\\
	\text{s.t.} \quad\; & {\overline {\bm Q}}_u \succeq \bm 0, ~ {\text {tr}} ({\overline {\bm Q}}_u) \leq P_u, ~\forall u \in {\cal U},
\end{align}
where ${\overline {\bm Q}}_u \in {\mathbb C}^{{\hat N}_u \times {\hat N}_u}$, ${\overline {\bm Q}}_{\cal U} = \{{\overline {\bm Q}}_1, \dots, {\overline {\bm Q}}_U\}$, and $\bm w_{\cal U}'' = \{\bm w_1'', \dots, \bm w_U''\}$.
Note that ${\overline C} ({\overline {\bm Q}}_{\cal U}, \bm w_{\cal U}'')$ has a similar definition as $C (\bm Q_{\cal U}, \bm w_{\cal U})$ in (\ref{capacity}).
The only difference lies in that the dimension at the user-side has changed from $N_u$ to ${\hat N}_u$.
Let ${\overline {\bm Q}}_{\cal U}''$ denote the optimal solution of (\ref{problem6}).
Then, it is obvious that ${\overline C} ({\overline {\bm Q}}_{\cal U}'', \bm w_{\cal U}'')$ is an upper bound to $C (\bm Q_{\cal U}^*, \bm w_{\cal U}^*)$, i.e.,
\begin{equation}\label{ub}
	{\overline C} ({\overline {\bm Q}}_{\cal U}'', \bm w_{\cal U}'') \geq C (\bm Q_{\cal U}^*, \bm w_{\cal U}^*).
\end{equation}
In addition, we can prove by following similarly steps as above that if $N_u, \forall u \in {\cal U}$ are all large, ${\overline C} ({\overline {\bm Q}}_{\cal U}'', \bm w_{\cal U}'')$ is approximate to $C^{\rm approx} ({\hat {\bm Q}}_{\cal U})$, i.e., 
\begin{equation}\label{ub_approx}
	{\overline C} ({\overline {\bm Q}}_{\cal U}'', \bm w_{\cal U}'') \approx C^{\rm approx} ({\hat {\bm Q}}_{\cal U}^*).
\end{equation}
Combining (\ref{lb}), (\ref{lb_approx}), (\ref{ub}), and (\ref{ub_approx}), if $N_u, \forall u \in {\cal U}$ are all large, (\ref{appro1}) is true. Theorem~\ref{theorem3} is thus proven.

\section{Proof of Theorem~\ref{lemma1}}\label{prove_lemma1}
When $M$ is large, $C (w)$ in (\ref{capacity_su}) can be approximated as
\begin{align}\label{approx}
& C (w) = \log \left| P \bm g (w) \bm g^H (w) + \bm I_M \right| \nonumber\\
& \overset{\text {(a)}}{=} \log \left( P \bm g^H (w) \bm g (w) + 1 \right) \nonumber\\
& \overset{\text {(b)}}{\approx} \log \!\left(\! P M \!\left[ a_{\text T} (\theta_1, w), \dots, a_{\text T} (\theta_L, w) \right]\! \bm \varGamma^H  \bm \varGamma \!\!\begin{bmatrix}\! a_{\text T}^* (\theta_1, w)\\ \vdots \\ a_{\text T}^* (\theta_L, w) \!\end{bmatrix}\! \!+\! 1 \!\right)	\nonumber\\
& = \log \left( \sum_{l=1}^L P M | a_{\text T} (\theta_l, w) \gamma_l|^2 + 1 \right) \nonumber\\
& = \log \left( \sum_{l=1}^L P M |\gamma_l|^2 + 1 \right)
\triangleq C_0, ~ \forall w \in [0, W],
\end{align}
where (a) holds due to (\ref{O1O2_1}), and (b) follows from using (\ref{g_w}) and the fact that if $M$ is large, $\bm A_{\text R}^H \bm A_{\text R} \approx \bm I_L$, which can be analogously proven as (\ref{approx1}).

\section{Proof of Theorem~\ref{theorem1}}\label{prove_theorem1}
We prove Theorem~\ref{theorem1} by separately discussing three different cases of $\psi_2^{\text {Im}}$, i.e., $\psi_2^{\text {Im}} = 0$, $\psi_2^{\text {Im}} > 0$, and $\psi_2^{\text {Im}} < 0$. First, if $\psi_2^{\text {Im}} = 0$, the objective function of (\ref{problem4}) satisfies
\begin{align}\label{cos_rhow}
\psi_2^{\text {Im}} \sin (\rho w) + \psi_2^{\text {Re}} \cos (\rho w) = \psi_2^{\text {Re}} \cos (|\rho| w) \leq |\psi_2^{\text {Re}}|,
\end{align}
which holds with equality if $\cos (|\rho| w)$ equals $- 1$ or $1$, depending on the sign of $\psi_2^{\text {Re}}$. If $\psi_2^{\text {Re}} \geq 0$, the upper bound in (\ref{cos_rhow}) can be achieved when $w = 0$. Hence,
\begin{equation}\label{opt_w3}
w^* = 0.
\end{equation}
Note that since $\cos (|\rho| w)$ is a periodic function, there may be multiple $w \in [0, W]$ that can make (\ref{cos_rhow}) hold with equality. Here we provide only one solution. If $\psi_2^{\text {Re}} < 0$, the minimum positive $w$ that satisfies $\cos (|\rho| w) = -1$ and thus makes (\ref{cos_rhow}) hold with equality is $\pi/ |\rho|$. Noticing that $w$ is in the interval $[0, W]$. Therefore, if $\pi/ |\rho| \in [0, W]$, 
\begin{equation}\label{opt_w4}
w^* = \pi/ |\rho|.
\end{equation}
Otherwise, (\ref{cos_rhow}) is always strict, and it can be easily checked that $\psi_2^{\text {Re}} \cos (|\rho| w)$ increases with $w$ in $[0, W]$. Therefore, 
\begin{equation}\label{opt_w5}
w^* = W.
\end{equation}
Combining (\ref{opt_w3}), (\ref{opt_w4}), and (\ref{opt_w5}), we obtain (\ref{opt_w1}).

Next, if $\psi_2^{\text {Im}} > 0$, using the auxiliary angle method of trigonometric functions, the objective function of (\ref{problem4}) can be rewritten and upper bounded as
\begin{align}\label{auxi_ang}
\psi_2^{\text {Im}} \sin (\rho w) \!+\! \psi_2^{\text {Re}} \cos (\rho w) & =\! \sqrt{(\psi_2^{\text {Im}})^2 \!+\! (\psi_2^{\text {Re}})^2} \sin \left( \rho w \!+\! \mu \right) \nonumber\\
& =\! |\psi_2| \sin \left( \rho w + \mu \right) \leq |\psi_2|,\!
\end{align}
where $\mu$ has been defined in (\ref{w_hat}). When $\rho > 0$, since $\mu \in [-\pi/2, \pi/2]$, the minimum positive $w$ that makes (\ref{auxi_ang}) hold with equality is $\frac{\pi/2 - \mu}{\rho}$. Considering that $w$ should be in the interval $[0, W]$, if $\frac{\pi/2 - \mu}{\rho} \in [0, W]$, 
\begin{equation}\label{solution_case1}
w^* = \frac{\pi/2 - \mu}{\rho}.
\end{equation}
Otherwise, if $\frac{\pi/2 - \mu}{\rho} \notin [0, W]$, inequality (\ref{auxi_ang}) is always strict, and it can be easily checked that the $w$ that maximizes $|\psi_2| \sin \left( \rho w + \mu \right)$ or $C(w)$ must be one of the boundary points of the interval $[0, W]$. Therefore,
\begin{equation}\label{opt_w7}
w^* = \left\{\!\!\!
\begin{array}{ll}
0, ~{\text {if}}~ \frac{\pi/2 - \mu}{\rho} \notin [0, W] ~{\text {and}}~ C(0) \geq C(W), \vspace{0.3em}\\
W, ~{\text {otherwise}}.
\end{array} \right.
\end{equation}
If $\rho < 0$, it can be proven similarly that
\begin{equation}\label{opt_w8}
w^* = \left\{\!\!\!
\begin{array}{ll}
\frac{- \pi/2 - \mu}{\rho}, ~{\text {if}}~ \frac{- \pi/2 - \mu}{\rho} \in [0, W], \vspace{0.3em}\\ 
0, ~{\text {if}}~ \frac{- \pi/2 - \mu}{\rho} \notin [0, W] ~{\text {and}}~ C(0) \geq C(W), \vspace{0.3em}\\
W, ~{\text {otherwise}}.
\end{array} \right.
\end{equation}
The details are omitted here for brevity.

Last, if $\psi_2^{\text {Im}} < 0$, the auxiliary angle method of trigonometric functions can still be applied to get the following bound
\begin{align}\label{auxi_ang2}
\psi_2^{\text {Im}} \sin (\rho w) \!+\! \psi_2^{\text {Re}} \cos (\rho w) \!=\! - |\psi_2| \sin \left( \rho w \!+\! \mu \right)  \leq |\psi_2|.\!\!
\end{align}
By following similar steps, we can prove that if $\rho > 0$,
\begin{equation}\label{opt_w9}
w^* = \left\{\!\!\!
\begin{array}{ll}
\frac{3 \pi/2 - \mu}{\rho}, ~{\text {if}}~ \frac{3 \pi/2 - \mu}{\rho} \in [0, W], \vspace{0.3em}\\ 
0, ~{\text {if}}~ \frac{3 \pi/2 - \mu}{\rho} \notin [0, W] ~{\text {and}}~ C(0) \geq C(W), \vspace{0.3em}\\
W, ~{\text {otherwise}},
\end{array} \right.
\end{equation}
and if $\rho < 0$,
\begin{equation}\label{opt_w10}
w^* = \left\{\!\!\!
\begin{array}{ll}
\frac{- 3 \pi/2 - \mu}{\rho}, ~{\text {if}}~ \frac{- 3 \pi/2 - \mu}{\rho} \in [0, W], \vspace{0.3em}\\ 
0, ~{\text {if}}~ \frac{- 3 \pi/2 - \mu}{\rho} \notin [0, W] ~{\text {and}}~ C(0) \geq C(W), \vspace{0.3em}\\
W, ~{\text {otherwise}}.
\end{array} \right.
\end{equation}
Then, we know that (\ref{opt_w2}) is true. This completes the proof.

\vspace{-2mm}
\section{Proof of Theorem~\ref{theorem2}}\label{prove_theorem2}
We prove (\ref{eq1}) by contrapositive. If (\ref{eq1}) is not true, there exist two possible cases, i.e., 
\begin{align}\label{eq1_1}
\log\! | \bm G (\bm w^*) \bm Q^* \bm G^H \!(\bm w^*) \!+\! \bm I_M |
\!>\! \log\! | \bm G^H \!(\bm w^\divideontimes) \bm F^\divideontimes \!\bm G (\bm w^\divideontimes) \!+\! \bm I_N |,
\end{align}
and 
\begin{align}\label{eq1_2}
\log\! | \bm G (\bm w^*) \bm Q^* \bm G^H \!(\bm w^*) \!+\! \bm I_M | \!<\! \log\! | \bm G^H \!(\bm w^\divideontimes) \bm F^\divideontimes \!\bm G (\bm w^\divideontimes) \!+\! \bm I_N |.
\end{align}
If (\ref{eq1_1}) holds, for given $\bm w^*$, the optimal solution of (\ref{problem16_Qw1}) must be $\bm Q^*$ since otherwise the assumption that $(\bm Q^*, \bm w^*)$ is the optimal solution of (\ref{problem16_Qw}) will be violated. In addition, let $\bm F^*$ be the optimal solution of (\ref{problem16_1}) for given $\bm w^*$. According to (\ref{capacity3}), we know that given $\bm w^*$, the optimal objective function values of (\ref{problem16_Qw1}) and (\ref{problem16_1}) are the same, i.e.,
\begin{align}\label{eq1_3}
& \log | \bm G (\bm w^*) \bm Q^* \bm G^H (\bm w^*) + \bm I_M | \nonumber\\
= & \log | \bm G^H (\bm w^*) \bm F^* \bm G (\bm w^*) + \bm I_N |.
\end{align}
Then, based on (\ref{eq1_1}) and (\ref{eq1_3}), we have
\begin{align}\label{eq1_4}
& \log | \bm G^H (\bm w^*) \bm F^* \bm G (\bm w^*) + \bm I_N | \nonumber\\
> & \log | \bm G^H (\bm w^\divideontimes) \bm F^\divideontimes \bm G (\bm w^\divideontimes) + \bm I_N |,
\end{align}
which contradicts the assumption that $(\bm F^\divideontimes, \bm w^\divideontimes)$ is the optimal solution of (\ref{problem16}). Similarly, it can be proven that if (\ref{eq1_2}) holds, the assumption that $(\bm Q^*, \bm w^*)$ is the optimal solution of (\ref{problem16_Qw}) will be violated. The equation (\ref{eq1}) is thus true.

Combining (\ref{eq1}) and (\ref{eq1_3}), we know that (\ref{eq2}) is true. It can be proven in an analogous way that (\ref{eq3}) is also true. This completes the proof of Theorem~\ref{theorem2}.

\bibliographystyle{IEEEtran}

\end{document}